# A reference Earth model for the heat producing elements and associated geoneutrino flux

**Yu Huang**
*Department of Geology, University of Maryland, College Park, MD 20742, USA*
*(yuhuang@umd.edu)*
**Viacheslav Chubakov and Fabio Mantovani**
*Dipartimento di Fisica, Università degli Studi di Ferrara, Ferrara, Italy*
*Istituto Nazionale di Fisica Nucleare, Sezione di Ferrara, Ferrara, Italy*
**Roberta L. Rudnick and William F. McDonough**
*Department of Geology, University of Maryland, College Park, MD 20742, USA*

The recent geoneutrino experimental results from KamLAND and Borexino detectors reveal the usefulness of analyzing the Earth's geoneutrino flux, as it provides a constraint on the strength of the radiogenic heat power and this, in turn, provides a test of compositional models of the bulk silicate Earth (BSE). This flux is dependent on the amount and distribution of heat producing elements (HPEs: U, Th and K) in the Earth's interior. We have developed a geophysically-based, three-dimensional global reference model for the abundances and distributions of HPEs in the BSE. The structure and composition of the outermost portion of the Earth, the crust and underlying lithospheric mantle, is detailed in the reference model, this portion of the Earth has the greatest influence on the geoneutrino fluxes. The reference model combines three existing geophysical models of the global crust and yields an average crustal thickness of 34.4±4.1 km in the continents and 8.0±2.7 km in the oceans, and a total mass (in $10^{22}$ kg) of oceanic, continental and bulk crust is 0.67±0.23, 2.06±0.25 and 2.73±0.48, respectively. *In situ* seismic velocity provided by CRUST 2.0 allows us to estimate the average composition of the deep continental crust by using new and updated compositional databases for amphibolite and granulite facies rocks in combination with laboratory ultrasonic velocities measurements. An updated xenolithic peridotite database is used to represent the average composition of continental lithospheric mantle. Monte Carlo simulation is used to predict the geoneutrino flux at 16 selected locations and to track the asymmetrical uncertainties of radiogenic heat power due to the log-normal distributions of HPE concentrations in crustal rocks.

## 1. Introduction

Determining the Earth's heat budget and heat production is critical for understanding plate tectonics and the thermal evolution of the Earth. Recent detection of geoneutrinos (electron anti-neutrinos generated during beta decay) offers a means to determine the U and Th concentrations in the Earth that is complementary to traditional cosmochemical or geochemical arguments [*Dye*, 2010]. However, since all three existing geoneutrino detectors are currently located within the continental crust (two in operation, another coming on line in 2013), the crustal contribution, which dominates the geoneutrino signal, must be subtracted in order to determine the signal from the mantle and core [*Dye*, 2012; *Fiorentini et al.*, 2012; *Šrámek et al.*, 2013].

Here we develop a three-dimensional global reference model that describes the inventory and distribution of the heat producing elements (HPEs: U, Th and K) in the bulk silicate Earth (BSE), along with uncertainties. The greatest resolution of the model resides in the outermost portions of the Earth – the crust and underlying lithospheric mantle, from whence the largest portion of the surface flux originates. The model is open-source and represents the first step in an effort to develop community ownership.

**1.1 Heat Producing Elements and Earth Differentiation**

Radioactivities of U, Th and K contribute about 99%, with a relative contribution of approximately 2:2:1, of the total radiogenic heat power of the Earth. Although the heat production rate for unit mass of Rb at natural isotopic abundance is higher than K, the contribution of Rb to the total radiogenic heat power is expected to be less than 1% [*Fiorentini et al.*, 2007], given the relative decay rates, and a K/Rb ratio of ~400 in the BSE [*McDonough and Sun*, 1995]. The other elements, such as La, Sm, etc., make negligible contributions to the total radiogenic power.

Uranium and Th are refractory lithophile elements, while K is a volatile lithophile element. The lithophile classification means that HPEs are expected to reside in the rocky portion of the Earth [*Goldschmidt*, 1933], though some have speculated that U and K may become slightly siderophile or chalcophile at high temperatures and pressures, and thus, may enter the Earth's core [e.g., *Lewis*, 1971; *Murrell and Burnett*, 1986; *Murthy et al.*, 2003]. The refractory nature of U and Th means that the Earth should have accreted with the full solar complement of these elements, whereas the volatility of K has led to its depletion in the Earth relative to the Sun and primitive chondritic meteorites [e.g., *McDonough*, 2003]. Thus, the concentration of K in the Earth is inferred from analyses of geological samples and its behavior relative to refractory elements.

Uranium, Th and K are all highly incompatible elements (defined as having crystal/melt partition coefficients much less than one), and, thus, are concentrated in melts relative to residues during partial melting. The Earth has experienced irreversible differentiation via melting and the ascent of these melts towards the surface, leading to the concentration of these elements in the outermost-layers of the planet. Thus, although the continental crust comprises only ~0.5% of the mass of the BSE, it contributes almost one third of the total radiogenic heat power, and refining the composition of the continental crust is an essential prerequisite to using geoneutrinos to “see” into the deeper levels of the Earth.

Compositional models for the BSE vary by nearly a factor of three in their U content (i.e., ~10 ng/g [*Javoy et al.*, 2010; *O'Neill and Palme*, 2008], ~20 ng/g [*Allègre et al.*, 1995; *Hart and Zindler*, 1986; *Lyubetskaya and Korenaga*, 2007a; b; *McDonough and Sun*, 1995; *Palme and O'Neill*, 2003], and ~30 ng/g [*Anderson*, 2007; *Turcotte and Schubert*, 2002; *Turcotte et al.*, 2001]). These models generally agree on a Th/U of 3.9 and a K/U of 14,000 [*Arevalo et al.*, 2009]. Compositional models for the continental crust (see summary in *Rudnick and Gao* [2003]) predict a U content of 1100 to 2700 ng/g, implying that anywhere between 30 and 45% of the budget of HPEs is stored in this thin skin of crust and that it is more than ~100-fold enriched over the modern mantle (i.e., ~13 ng/g of U), assuming a geochemical model for the BSE of *McDonough and Sun* [1995]. Geoneutrino data, when available for several sites on the Earth, should be able to define permissible models for the BSE and the continental crust.

### 1.2 Geoneutrinos

The Earth is an electron antineutrino star that emits these nearly massless particles at a rate of ~$10^6$ $cm^{-2}$ $s^{-1}$ [e.g., *Enomoto et al.*, 2007; *Fiorentini et al.*, 2007; *Kobayashi and Fukao*, 1991; *Mantovani et al.*, 2004]. Geoneutrinos are electron antineutrinos produced within the Earth by beta-minus decay when a neutron decays to a proton via the weak interaction. This decay process, in which a down quark transforms to an up quark, is mediated by the emission of a $W^-$ boson along with an electron, and a charge neutral electron anti-neutrino. Because of their vanishingly small cross-section for interaction, ~$10^{-44}$ $cm^2$, matter is virtually transparent to these particles and they have about a 50% chance of passing through a light-year of lead without interaction. By comparison, the fusion processes inside the core of the Sun produce neutrinos, the anti-matter lepton counterpart of antineutrinos, which bathe the Earth's surface with a flux that is ~$10^4$ greater than the geoneutrino flux [*Bahcall et al.*, 2005]. The term geoneutrino distinguishes natural emissions of electron antineutrinos from those radiated from nuclear reactors.

To date, geoneutrino flux measurements have been made at two detectors, KamLAND, at the Kamioka mine in Japan [*Araki et al.*, 2005; *Gando et al.*, 2011], and Borexino, at the Gran Sasso underground laboratories in Italy [*Bellini et al.*, 2010], and provide constraints on the quantities of U and Th inside the Earth. The SNO+ detector at the Sudbury Neutrino Observatory, Canada [*Chen*, 2006], will come on-line in 2013 and will deliver significant new data on the geoneutrino flux from the Archean Superior Craton and surrounding North American plate.

Geoneutrinos originating from U and Th can be distinguished based on their energy spectra, e.g., geoneutrinos with E > 2.25 MeV are produced only in the $^{238}$U chain [e.g., *Araki et al.*, 2005]. Liquid scintillator detectors work by sensing light generated during antineutrino-proton interactions: $\overline{\nu}_e + p \rightarrow e^+ + n$, when the $\overline{\nu}_e$ has ≥1.806 MeV energy, which is the energy needed to transform the proton, p, to a positron $e^+$ and a neutron, n. Of the total geoneutrino flux, only small portions of antineutrinos generated in the $^{238}$U and $^{232}$Th decay chains can be detected by this mechanism. The hydrogen nuclei, which are in abundant supply in hydrocarbon ($C_nH_{2n}$) based liquid scintillator detectors, act as the target for transiting antineutrinos. The directionality of antineutrinos is presently undetectable and, thus, the detectors are sensitive only to the integrated flux. Fortunately, because the geoneutrino flux at a detector decreases with distance from the source via the inverse square law, geoneutrinos can be used to detect regional differences in the distribution of U and Th in the continents, and, in principle, large-scale features in the mantle [*Dye*, 2010; *Šrámek et al.*, 2013]. Thus, accurate and precise detection of the surface flux of geoneutrinos, coupled with geochemical and geophysical models of local and global crust, will enable quantitative tests of compositional models of the planet.

## 1.3 Modeling the Earth's Heat Producing Elements

We can model the Earth's geoneutrino flux by assigning physical and chemical data to a set of spatially defined voxels (a volume element, comparable to a three-dimensional pixel). Such a model can be compared to surface heat flow measurements and various mass balance models for the composition of the Earth and its internal reservoirs (i.e., crust, mantle, and core). Towards this goal, an enormous amount of geophysical and geochemical data have been collected and shared on line in the past few decades. This information can be integrated into a broader framework in order to evaluate the nature and or existence of planetary features, such as chemical compositions of thermochemical piles in the mantle [*Šrámek et al.*, 2013], the characteristics of a residual layer from a basal magma ocean [*Labrosse et al.*, 2007; *Lee et al.*, 2010], and/or the presence of an early Earth enriched reservoir that was sequestered at the core-mantle boundary [*Boyet and Carlson*, 2005]. Future geoneutrino observations will bring clarity to the debates regarding the mantle Urey ratio (the ratio of radiogenic heat in the mantle to total mantle heat flux) and the forces driving plate tectonics and mantle convection [e.g., *Korenaga*, 2008; *Labrosse and Jaupart*, 2007]. These data will also define aspects of the Earth's thermal evolution.

To build the reference crustal model, we combine (1) geophysical information from seismic refraction measurements [*Bassin et al.*, 2000; *Laske and Masters*, 1997], surface wave dispersion data [*Shapiro and Ritzwoller*, 2002] and gravity anomalies observations [*Negretti et al.*, 2012; *Reguzzoni and Tselfes*, 2009], (2) estimates of the average compositions of the upper continental crust [*Rudnick and Gao*, 2003], global sediments [*Plank*, 2013] and oceanic crust [*White and Klein*, 2013], (3) laboratory ultrasonic measurements of deep-crustal rock types, and (4) new and updated geochemical compilations for deep crustal rocks and lithospheric peridotites to provide new insights on the composition of the deep crust and continental lithospheric mantle (CLM). In order to make more accurate predictions of the geoneutrino flux at current detectors and possible future detector sites, we define the mass and geometry of continental crust, quantify the amount and distribution of the HPEs and characterize their lateral and vertical variations in the crust. We also provide uncertainties for all estimates. For the first time, the geoneutrino flux originating from the CLM is estimated. Collectively, this model allows the geoneutrino flux from the deep Earth to be defined more accurately, given that a large proportion of total signal at any given detector located in the continental crust is derived from this thin outer crustal layer.

## 2. Methodology and Reference States

Here we describe the Earth as the sum of its metallic, silicate, and hydrospheric shells. The silicate shell of the Earth (equivalent to the BSE) is considered to be the main repository of HPEs, and we focus on understanding internal differentiation of this region (Fig. 1). The BSE is composed of five dominant domains, or reservoirs: the DM (Depleted Mantle, which is the source of mid-ocean-ridge-basalts -- MORB), the EM (Enriched Mantle, which is the source of Ocean Island Basalts -- OIB), the CC (continental crust), the OC (oceanic crust), and the lithospheric mantle (LM). It follows that BSE = DM + EM + CC + OC + LM. The modern convecting mantle is composed of the DM and the EM. We do not include a term for a hidden reservoir, which may or may not exist in the BSE; its potential existence is not a consideration of this paper.

### 2.1. Selection of Flux Calculation Sites

Although geoneutrinos can be measured, in principle, anywhere on the Earth, the experiments need to be carried out in underground (or underwater) laboratories in order to shield detectors from cosmic radiation; only a few locations therefore have particular experimental interest. We have calculated the fluxes at 16 sites where the exploration of the Earth through geoneutrinos is either currently underway (Kamioka, Japan, with the KamLAND experiment [*Araki et al.*, 2005; *Gando et al.*, 2011]; Gran Sasso, Italy, with the Borexino experiment [*Alvarez Sanchez et al.*, 2012]; Sudbury, Ontario, Canada, with the SNO+ experiment [*Chen*, 2006]), or where such experiments have been proposed or could be planned (Table 1). Hawaii (Hanohano [*Dye*, 2010]), Baksan (Baksan Neutrino Observatory [*Buklerskii et al.*, 1995]), Homestake (Deep Underground Science and Engineering Laboratory [*Tolich et al.*, 2006]), Curacao (Earth AntineutRino TomograpHy [*De Meijer et al.*, 2006]) and Daya Bay (Daya Bay II [*Wang*, 2011]) are all sites that have been proposed for constructing liquid scintillator detectors capable of detecting geoneutrinos. LAGUNA (Large Apparatus studying Grand Unification and Neutrino Astrophysics) is looking for the best site in Europe where the LENA (Low Energy Neutrino Astronomy) experiment [*Wurm et al.*, 2012] could be built: seven prospective underground sites in Europe (Pyhasalmi, Boulby, Canfranc, Fréjus, Slanic and SUNLAB (see LAGUNA website)) are being investigated. Finally, we also include the sites where the maximum and minimum geoneutrinos signal on the Earth's surface is expected: the Himalaya and Rurutu Island (Pacific Ocean), respectively.

**2.2. Structure and Mass of the Crust**

In 1998, the CRUST 5.1 model [*Mooney et al.*, 1998] was published as a refinement of the previous 3SMAC model [*Nataf and Richard*, 1996]. The model included 2592 voxels on a 5°×5° grid, and reported the thickness and physical properties of all ice and sediment accumulations and of normal and anomalous oceanic crust. Vast continental regions (large portions of Africa, South America, Antarctica and Greenland) lacked direct observations, and the predictions for these areas were obtained by extrapolation based on the crustal structure. Taking advantage of a compilation of new reflection and refraction seismic data, a global crustal model at 2°×2° resolution (CRUST 2.0) by *Bassin et al.* [2000] provided an update to CRUST 5.1. This model incorporates 16,200 crustal voxels and 360 key profiles that contain the thickness, density and velocity of compressional (Vp) and shear waves (Vs) for seven layers (ice, water, soft sediments, hard sediments, upper, middle and lower crust) in each voxel. The Vp values are based on field measurements, while Vs and density are estimated by using empirical Vp-Vs and Vp-density relationships, respectively [*Mooney et al.*, 1998]. For regions lacking field measurements, the seismic velocity structure of the crust is extrapolated from the average crustal structure for regions with similar crustal age and tectonic setting [*Bassin et al.*, 2000]. Topography and bathymetry are adopted from a standard database (ETOPO-5). The same physical and elastic parameters are reported in a global sediment map digitized on a 1°×1° grid [*Laske and Masters*, 1997]. The accuracies of these models are not specified and they must vary with location and data coverage.

The crust in our reference Earth model is composed of 64,800 voxels at a resolution of 1°×1°, and is divided into two main reservoirs: oceanic crust (OC) and continental crust (CC). In the OC we include the oceanic plateaus and the melt-affected oceanic crust of *Bassin et al.* [2000]. The other crustal types identified in CRUST 2.0 are considered to be CC, including oceanic plateaus comprised of continental crust (the so-called "W" tiles of *Bassin et al.* [2000]), which are mainly found in the north of the Scotia Plate, in the Seychelles Plate, in the plateaus around New Zealand (Campbell Plateau, Challenger Plateau, Lord Howe Rise and Chatham Rise), and on the northwest European continental shelf. For each voxel, we adopt the physical information (density and relative thickness) of three sediment layers from the global sediment map [*Laske and Masters*, 1997]; for upper, middle and lower crust we adopt the physical and elastic parameters (Vp and Vs) from CRUST 2.0 [*Bassin et al.*, 2000].

Evaluation of the uncertainties of the crustal structure is complex, as the physical parameters (thickness, density, Vp and Vs) are correlated, and their direct measurements are inhomogeneous over the globe [*Mooney et al.*, 1998]. Seismic velocities generally have smaller relative uncertainties than thickness [*Christensen and Mooney*, 1995], since seismic velocities (Vp) are measured directly in the refraction method, while the depths of refracting horizons are successively calculated from the uppermost to the deepest layer measured. The uncertainties of seismic velocities in some previous global crustal models were estimated to be 3-4% [*Holbrook et al.*, 1992; *Mooney et al.*, 1998]. To be conservative, we adopt 5% (1-sigma) uncertainties for both Vp and Vs in our reference crustal model.

The accuracy of the crustal thickness model is crucial to our calculations, as the uncertainties of all boundary depths affect the global crustal mass, the radiogenic heat power and the geoneutrino flux. In particular, uncertainties in Moho depths are a major source of uncertainty in the global crustal model. Although CRUST 2.0 does not provide uncertainties for global crustal thickness, the previous 3SMAC topographic model [*Nataf and Richard*, 1996] included the analysis of crust-mantle boundary developed by *Čadek and Martinec* [1991], in which the average uncertainties of continental and oceanic crustal thickness are 5 km and 3 km (1-sigma), respectively. Fig. 2a shows the dispersion of the thickness of all CC voxels in CRUST 2.0. The surface area weighted average continental and oceanic crustal thickness (ice and water excluded, sediment included) in CRUST 2.0 is 35.7 km and 7.5 km, respectively.

Gravity data can be used to constrain the crustal thickness and is especially important in areas that lack seismic observations and crustal density [*Mooney et al.*, 1998; *Tenzer et al.*, 2009]. The GOCE satellite (Gravity field and steady-state Ocean Circulation Explorer), launched in March, 2009, is the first gravity gradiometry satellite mission dedicated to providing an accurate and detailed global model of the Earth's gravity field with a resolution of about 80 km and an accuracy of 1-2 cm in terms of geoid [*Pail et al.*, 2011]. The GEMMA project (GOCE Exploitation for Moho Modeling and Applications) has developed the first global high-resolution map (0.5°×0.5°) of Moho depth by applying regularized spherical harmonic inversion to gravity field data collected by GOCE and preprocessed using the space-wise approach [*Reguzzoni and Tselfes*, 2009; *Reguzzoni and Sampietro*, 2012]. This global crustal model is obtained by dividing the crust into different geological provinces and defining a characteristic density profile for each of them. Using the database of GEMMA [*Negretti et al.*, 2012], we calculate the surface area weighted average thicknesses of CC and OC to be 32.7 km and 8.8 km, respectively (Fig. 2b).

Another way to evaluate the global crustal thickness is by utilizing the phase and group velocity measurements of the fundamental mode of Rayleigh and Love waves. *Shapiro and Ritzwoller* [2002] used a Monte Carlo method to invert surface wave dispersion data for a global shear-velocity model of the crust and upper mantle on a 2°×2° grid (CUB 2.0), with *a priori* constraints (including density) from the CRUST 5.1 model [*Mooney et al.*, 1998]. With the dataset of this model (courtesy of V. Lekic), the surface area weighted average thicknesses of the CC and OC are 34.8 km and 7.6 km, respectively (Fig. 2c).

The three global crustal models described above were obtained by different approaches and the constraints on the models are slightly dependent. Ideally, the best solution for a geophysical global crustal model is to combine data from different approaches: reflection and refraction seismic body wave, surface wave dispersion, and gravimetric anomalies. In our reference model the thickness and its associated uncertainty of each 1°×1° crustal voxel is obtained as the mean and the half-range of the three models. The surface area weighted average thicknesses of CC and OC are 34.4 ±4.1 km (Fig. 2d) and 8.0±2.7 km (1-sigma) for our reference crustal model, respectively. The uncertainties reported here are not based on the dispersions of thicknesses of CC and OC voxels, but are the surface area weighted average of uncertainties of each voxel's thickness. Our estimated average CC thickness is about 16% less than 41 km determined previously by *Christensen and Mooney* [1995] (see their Fig. 2) on the basis of available seismic refraction data at that time and assignment of crustal type sections for areas that were not sampled seismically. However, their compilation did not include continental margins, nor submerged continental platforms, which are included in the three global crustal models used here. Inclusion of these areas will make the CC thinner, on average, than that based solely on exposed continents.

Adopting from CRUST 2.0 the well-established thicknesses of water and ice, and the densities and relative proportions of each crustal layer, we calculate the masses of all crustal layers, including the bulk CC and OC (Table 3). Summing the masses of sediment, upper, middle and lower crust, the total masses of CC and OC are estimated to be $M_{CC} = (20.6\pm2.5) \times10^{21}$ kg and $M_{OC} = (6.7\pm2.3) \times10^{21}$ kg (1-sigma). Thus, the fractional mass contribution to the BSE of the CC is 0.51% and the contribution of the OC is 0.17%. The uncertainty of crustal thickness of each voxel is dependent on that of other voxels, but with undeterminable correlation, due to the fact that the three crustal models are mutually dependent, and the estimates of crustal thicknesses for some voxels are extrapolated from the others. Considering these complexities, we make the conservative assumption that the uncertainty of Moho depth in each voxel is totally dependent on that of all the others. Compared to the total crustal mass (i.e., CC + OC) derived directly from CRUST 2.0 ($27.9\times10^{21}$ kg) [*Dye*, 2010], the total crustal mass in our reference model ($(27.3\pm4.8) \times10^{21}$ kg) is ~2% lower, but within uncertainty. Although the CC covers only ~40% of the Earth's surface, it represents ~75% of the crustal mass; it is also the reservoir with the highest concentration of HPEs. Uncertainties in the concentrations of HPEs play a prominent role in constraining the crustal radiogenic heat power and geoneutrino flux, as discussed in Section 6.

### 2.3. The Lithospheric Mantle

Previous models of geoneutrino flux [*Dye*, 2010; *Enomoto et al.*, 2007; *Fogli et al.*, 2006; *Mantovani et al.*, 2004] have relied on CRUST 2.0 and the density profile of the mantle, as given by PREM (Preliminary Earth Reference Model, a 1-D seismologically based global model; [*Dziewonski and Anderson*, 1981]). In these models the crust and the mantle were treated as two separate geophysical and geochemical reservoirs. In particular, the mantle was conventionally described as a shell between the crust and the core, and considered compositionally homogeneous [*Dye*, 2010; *Enomoto et al.*, 2007]. These models didn't consider the heterogeneous topography of the base of the crust, or the likely differences in composition of the lithospheric mantle underlying the oceanic and continental crusts.

We treat the LM beneath the continents as a distinct geophysical and geochemical reservoir that is coupled to the crust in our reference Earth model (Fig. 1). We assume that the LM beneath the oceans is compositionally identical to DM, and therefore we make no attempt to constrain its thickness. The thickness of the CLM is variable under each crustal voxel, with the top corresponding to the Moho surface and the bottom being difficult to constrain [*Artemieva*, 2006; *Conrad and Lithgow-Bertelloni*, 2006; *Gung et al.*, 2003; *Pasyanos*, 2010]. The seismically, thermally and rheologically-defined depth to the base of the lithosphere may not be the same [*Jaupart and Mareschal*, 1999; *Jaupart et al.*, 1998; *Jordan*, 1975; *Rudnick and Nyblade*, 1999], and the thickness of the lithosphere can vary significantly across tectonic provinces, ranging from about 100 km in areas affected by Phanerozoic tectonism, to ≥250 km in stable cratonic regions [*Artemieva*, 2006; *Pasyanos*, 2010]. Here, we adopt 175±75 km (half-range uncertainty; 1-sigma) as representative of the average depth to the base of CLM.

The composition of the CLM is taken from an updated database of xenolithic peridotite compositions [*McDonough*, 1990] (Appendix D, DOI: 10.1594/IEDA/100247). The density profile of CLM under each crustal voxel is calculated using the linear parameterization described in PREM. The mass of CLM is reported in Table 3; the main source of uncertainty comes from the average depth of the base of CLM, while the uncertainty on Moho depth gives a negligible contribution.

### 2.4 The Sublithospheric Mantle

Deeper in the Earth, direct observations decrease dramatically, particularly, direct sampling of rocks for which geochemical data may be obtained. On the other hand, geoneutrinos are an extraordinary probe of the deep Earth. These particles carry to the surface information about the chemical composition of the whole planet and, in comparison with other emissions of the planet (e.g., heat or noble gases), they escape freely and instantaneously from the Earth's interior.

The structure of mantle between the base of lithosphere and the core-mantle boundary (CMB) has been a topic of great debate. Tomographic images of subducting slabs suggest deep mantle convection [e.g., *van der Hilst et al.*, 1997], while some geochemical observations favor a physically and chemically distinct upper and lower mantle, separated by the transition zone at the 660 km seismic discontinuity [e.g., *Kramers and Tolstikhin*, 1997; *Turcotte et al.*, 2001]. Within the geochemical community, there is considerable disagreement regarding the composition of the upper and lower mantle [*Allègre et al.*, 1996; *Boyet and Carlson*, 2005; *Javoy et al.*, 2010; *McDonough and Sun*, 1995; *Murakami et al.*, 2012].

Evaluation of the detailed structure of the mantle is not a priority of this paper, and in our model we divide the sublithospheric mantle into two reservoirs that are considered homogeneous. For simplicity, we assume these to be the depleted mantle (DM), which is on the top, and the underlying spherically symmetrical enriched mantle (EM) (Fig. 1). The DM is the source region for MORB, which provide constraints on its chemical composition [*Arevalo and McDonough*, 2010; *Arevalo et al.*, 2009]. The DM under CC and OC is variable in thickness due to the variable lithospheric thicknesses (Fig.1). The EM is an enriched reservoir beneath the DM, and the boundary between the two reservoirs, extending up to 710 km above the CMB, is estimated by assuming that EM accounts 18% of the total mass of the mantle [*Arevalo et al.*, 2009; *Arevalo et al.*, 2012]. The abundances of HPEs in the DM is ten times less than the global average MORB abundances [*Arevalo and McDonough*, 2010]; the enrichment factor of EM over DM is estimated through a mass balance of HPEs in the mantle, assuming a BSE composition of *McDonough and Sun* [1995]. The compositions of the DM and EM (without any associated uncertainties) are reported in Table 3. *Šrámek et al.* [2013] provide a detailed assessment of how different geophysical and geochemical mantle models influence the calculated geoneutrino fluxes from Earth's mantle.

The masses of DM and EM in our reference model (Table 3) are calculated by modeling the mantle density profile using the coefficients of the polynomials reported in PREM in spherical symmetry. The total mantle mass is well-known, based on the terrestrial moment of inertia and the density-depth profile of the Earth [*Yoder*, 1995]. The total mass of the mantle in our model (CLM+DM+EM) is $4.01 \times 10^{24}$ kg, in good agreement with the values reported by *Anderson* [2007] and *Yoder* [1995]. These results, combined with assumed abundances of HPEs in different reservoirs, will be used in the following sections to predict the geoneutrino flux and the global radiogenic heat power of the Earth.

## 3. Compositions of Earth Reservoirs

Here we review assumptions, definitions and uncertainties in modeling the structure and composition of all reservoirs in the reference model except for the deep CC and CLM, for which we derive new estimates based on several new and updated databases, as described in Section 4. First-order constraints on the Earth's structure are taken from PREM, and a model for the composition of the Earth [*McDonough*, 2003; *McDonough and Sun*, 1995]. Beyond that, we consider other input models and their associated uncertainties (Table 3).

### 3.1 The Core

Following the discussion in *McDonough* [2003], the Earth's core is considered to have negligible amounts of K, Th and U.

### 3.2 BSE Models and Uncertainties

A first step in determining the compositions of DM and EM in the reference model is to determine the composition of the BSE. Methods used to estimate the amount of K, Th and U in the BSE are principally based on cosmochemical, geochemical, and/or geodynamical data. Estimates based on U, a proxy for the total heat production in the planet, given planetary ratios of Th/U ~4 and K/U ~$10^4$, differ by almost a factor of three in the absolute HPE masses in the BSE, i.e., between $0.5\times10^{17}$ and $1.3\times10^{17}$ kg [*Šrámek et al.*, 2013].

A cosmochemical estimate for the BSE, which yields the lowest U concentration, matches the Earth's composition to a certain class of chondritic meteorites, the enstatite chondrites. *Javoy et al.* [2010] and *Warren* [2011] noted the similarity in chemical and isotopic composition between enstatite chondrites and the Earth. *Javoy et al.* [2010] constructed an Earth model from these chondritic building blocks and concluded that the BSE has a markedly low U content (i.e., 12 ng/g or $0.5\times10^{17}$ kg) and a total radiogenic heat production of 11 TW, using their preferred Th/U of 3.6 and K/U of 11,000. This model requires that the lower two thirds of the mantle is enriched in silica, has a markedly lower Mg/Si value and different mineralogical composition than that of the upper mantle [e.g., *Murakami et al.*, 2012], and that the bulk of the HPEs is concentrated in the CC. However, large scale, vertical differences in the upper and lower mantle composition are seemingly inconsistent with seismological evidence for subducting oceanic plates plunging into the deep mantle and stirring the entire convecting mantle.

A BSE model with similarly low HPEs was proposed by *O'Neill and Palme* [2008]. This model has only about 10 ng/g (i.e., $0.4\times10^{17}$ kg) of U based on the budget balance argument for the $^{142}$Nd and $^{4}$He flux, and it invokes the loss of up to half of the planetary budget of Th and U (and other highly incompatible elements) due to collisional erosion processes shortly following Earth accretion. The major concern with models that predict the BSE as having low overall HPE abundances is that this requires low radiogenic heat production in the mantle; the modern mantle is expected to have only ~3 ng/g of U and ~3 TW of radiogenic power, with the remaining fraction concentrated in the CC.

A geochemical method for modeling the BSE uses a combined approach of geochemical, petrologic and cosmochemical data to deconvolve the compositional data from the mantle and crustal samples [e.g., *McDonough and Sun*, 1995; *Palme and O'Neill*, 2003]. These models predict about ~$0.8\times10^{17}$ kg U (i.e., ~20 ng/g) in the BSE, have a relatively homogeneous major element composition throughout the mantle, and are consistent with elasticity models of the mantle and broader chondritic compositional models of the planet. Being based on samples, this method suffers from the fact that we may not sample the entire BSE and thus may not identify all components in the mantle.

The third approach to estimating the HPEs in the BSE is based on the surface heat flux, and derives solutions to the thermal evolution of the planet by examining the relative contributions of primordial heat and heat production needed to maintain a reasonable fit to the secular cooling record [e.g., *Anderson*, 2007; *Turcotte and Schubert*, 2002]; the compositions derived using this method are referred to here as geodynamical models. Such geodynamical models predict up to ~$1.2\times10^{17}$ kg U (~30 ng/g) in the BSE and require that more than 50% of the present heat flow is produced by radioactive decay. Defining the convective state of the mantle in terms of Rayleigh convection, these models compare the force balance between buoyancy and viscosity, versus that between thermal and momentum diffusivities, and conclude that conditions in the mantle greatly exceed the critical Rayleigh number for the body, which marks the onset of convection. These models, however, also require marked differences in the chemical and mineralogical composition of the upper and lower mantle, but differ from that of the cosmochemical models. A higher U content for the mantle translates into higher Ca and Al contents (i.e., higher clinopyroxene and garnet in the upper mantle or higher Ca-perovskite in the lower mantle), along with the rest of the refractory elements [*McDonough and Sun*, 1995], which, in turn, requires that the lower mantle has a higher basaltic component than envisaged for the upper mantle.

### 3.3 Sublithospheric Mantle (DM and EM)

Here we adopt the model of *McDonough and Sun* [1995] for the BSE, with updates for the absolute HPE contents given in *Arevalo et al.* [2009]. In addition, we use the definitions given by *Arevalo et al.* [2009] for the modern mantle, which is composed of two domains: a depleted mantle, DM, and a lower enriched mantle, EM. We envisage no gross compositional differences in major elements between the two domains, although the lowermost portion of the mantle is assumed to be the source for OIB magmas and is consequently enriched in incompatible elements (including HPEs) due to recycling of oceanic crust [*Hofmann and White*, 1983].

### 3.4 Continental Lithospheric Mantle (CLM)

The composition of the CLM adopted here stems from the earlier studies of *McDonough* [1990] and *Rudnick et al.* [1998], updated with newer literature data (see Section 4). As described above, the CLM is taken as the region below the Moho to 175±75 km depth under the CC. These limits are set arbitrarily to cover the full range of variation seen in different locations, ~100 km in orogens and extensional regions and reaching ~250 km beneath cratons, but it allows for the inclusion of a CLM that is likely to have a slight enrichment in HPEs due to secondary processes (e.g., mantle metasomatism). A future goal of related studies is the incorporation of gravimetric anomaly data and regional tomographic models, which may provide better geographical resolution regarding the depth to the lithosphere-asthenosphere boundary.

### 3.5 Crustal Components, Compositions and Uncertainties

Compositional estimates for some portions of the crust are adopted from previous work, whereas the composition of the deep continental crust is re-evaluated in Section 4.

3.5.1 Sediments: We adopt the average composition of sediments and reported uncertainties in the GLOSS II model (GLObal Subducting Sediments) [*Plank*, 2013].

3.5.2 Oceanic Crust: Areas in CRUST 2.0 labeled “A” and “B” are here considered oceanic crust. We assume an average oceanic crust composition as reported by *White and Klein* [2013], and adopt a conservative uncertainty of 30%. Seawater alteration can lead to enrichment of K and U in altered oceanic crust [*Staudigel*, 2003]. However, the oceanic crust makes negligible contributions to the geoneutrino flux and radiogenic heat power in the crust (Tables 2 and 3), and increasing the U and K contents in the oceanic crust by a factor of 1.6, as suggested by *Porter and White* [2009], has no influence on the outcomes of this study. We treat the three seismically defined layers of basaltic oceanic crust reported by *Mooney et al.* [1998] as having the same composition as average oceanic crust.

3.5.3 Upper Continental Crust: We adopt the compositional model reported by *Rudnick and Gao* [2003] for the upper continental crust and the uncertainties reported therein. Following *Mooney et al.* [1998], the upper continental crust is defined seismically as the uppermost crystalline region in CRUST 2.0, having an average Vp of between 5.3 and 6.5 km $s^{-1}$.

## 4. Refined Estimates for the Composition of the Deep Continental Crust and Continental Lithospheric Mantle

### 4.1 General Considerations

Given the large number of high-quality geochemical analyses now available for medium- to high-grade crustal metamorphic rocks, peridotites, ultrasonic laboratory velocity measurements and, especially, the large numbers of seismic refraction data for the crust (and their incorporation into CRUST 2.0), we re-evaluate here the composition of the deep CC and lithospheric mantle.

For the lithospheric mantle, we have updated the geochemical database for both massif and xenolithic peridotites of *McDonough* [1990] and *Rudnick et al.* [1998], as detailed in Section 4.3.2. For the deep CC, we follow the approach used by *Rudnick and Fountain* [1995] and *Christensen and Mooney* [1995], who linked laboratory ultrasonic velocity measurements to the geochemistry of various meta-igneous rocks. Laboratory measurements of Vp and Vs of both amphibolite and granulite facies rocks are negatively correlated with their $SiO_2$ contents (Fig. 3). This correlation allows one to estimate the bulk chemical composition of the lower and middle CC using seismic velocity data [*Christensen and Mooney*, 1995; *Rudnick and Fountain*, 1995].

*Behn and Kelemen* [2003], following *Sobolev and Babeyko* [1994], examined the relationship between Vp and major elements abundances of anhydrous igneous and meta-igneous rocks by making thermodynamic calculations of stable mineral assemblages for a variety of igneous rock compositions at deep crustal conditions, and then calculating their seismic velocities. They found a correlation between composition and seismic velocities, but also found very broad compositional bounds for a specific Vp in the deep CC, and concluded that P-wave velocities alone are insufficient to provide constraints on the deep crustal composition. In particular, they noted that *in situ* P-wave velocities in the lower crust of up to 7.0 km/s (corresponding to room temperature and 600 MPa Vp of 7.14 km/s calculated for an average crustal geotherm of 60 $mW/m^2$, using the temperature derivative given below) may reflect granulite-facies rocks having dacitic (~60 wt.% $SiO_2$) compositions. However, such broad compositional bounds are not observed in the laboratory data plotted in Fig. 3. For example, the $SiO_2$ content of rocks with Vp of ~7.1 km/s ranges from 42 to 52 wt.% $SiO_2$ for both amphibolite and granulite-facies lithologies.

We conclude that the correlation between seismic velocities and $SiO_2$, and the range in velocities at a given $SiO_2$ (Fig. 3), allow quantitative estimates of deep crustal composition and associated uncertainties. In the next three sections, we describe, in detail, the methodology employed here.

### 4.2 *In Situ* Velocity to Rock Type

Ultrasonic compressional and shear wave velocities have been determined for a variety of crustal rocks at different pressures and temperatures [e.g., *Birch*, 1960]. We have compiled published laboratory seismic velocity data for deep crustal rock types and summarize their average seismic properties at a confining pressure of 0.6 GPa and room temperature (Appendix A, DOI: 10.1594/IEDA/100238; Fig. 4; Table 4).

Several selection criteria are applied to the dataset. The compilation includes only data for grain-boundary-fluid free and unaltered rocks whose laboratory measurements were made in at least three orthogonal directions. We limit our compilation to measurements made at pressures ≥0.6 GPa in order to simulate pressures appropriate for the deep crust. Complete or near-complete closure of microcracks in the samples included in the compilation was ascertained by examining whether the seismic velocities increase linearly with pressure after reaching 0.4 GPa. Physical properties of xenoliths are usually significantly influenced by irreversible grain boundary alteration that occurs during entrainment [*Parsons et al.*, 1995; *Rudnick and Jackson*, 1995]. Since such alteration is not likely to be a feature of *in situ* deep crust, xenolith data are excluded from our compilation.

Metamorphosed igneous rocks are subdivided into felsic, intermediate, and mafic groups according to their $SiO_2$ contents, following the International Union of Geological Sciences (IUGS) classification of igneous rocks [*Le Bas and Streckeisen*, 1991] (i.e., $SiO_2$ = 45-52 wt.% for mafic, 52-63 wt.% for intermediate and >63 wt.% for felsic). Each group of meta-igneous samples is further subdivided into two sub-groups based on metamorphic facies and/or mineralogy: amphibolite facies and granulite facies, which are taken to represent the main rock types in the middle and lower CC, respectively. Amphibolite facies meta-igneous rocks normally contain no orthopyroxene, while granulite facies rocks contain orthopyroxene and/or clinopyroxene. Pelitic rocks (metamorphosed shales) have also been subdivided into amphibolite facies and granulite facies groups: muscovite and biotite are abundant phases in amphibolite facies metapelite and absent or minor phases in granulite facies metapelite. In some cases we revised the published classification of samples based on the reported mineralogy and/or chemical composition in order to be consistent with the classifications described above. The frequency distributions of Vp and Vs are generally Gaussian for the different deep crustal rock types (Fig. 4); we therefore adopt the mean and 1-sigma standard deviation as being representative of a given population (Table 4, Fig. 3).

Because seismic velocities of rocks in the deep crust are strongly influenced by pressure and temperature, we correct the compiled laboratory-measured velocities for all rock groups (which were attained at 0.6 GPa and room temperature) to seismic velocities appropriate for pressure-temperature conditions in the deep crust. To compare our compiled laboratory ultrasonic velocities to the velocities in the crustal reference model, we apply pressure and temperature derivatives of 2 $\times 10^{-4}$ km s$^{-1}$ MPa$^{-1}$ and -4 $\times 10^{-4}$ km s$^{-1}$ $^{o}$C$^{-1}$, respectively, for both Vp and Vs [*Christensen and Mooney*, 1995; *Rudnick and Fountain*, 1995], and assume a typical conductive geotherm equivalent to a surface heat flow of 60 mW·m$^{-2}$ [*Pollack and Chapman*, 1977]. Using the *in situ* Vp and Vs profiles for the middle (or lower) CC of each voxel given in CRUST 2.0, we estimate the fractions of felsic and mafic amphibolite facies (or granulite facies) rocks by comparing the *in situ* seismic velocities with the temperature- and pressure-corrected laboratory-measured velocities under the assumption that the middle (or lower) CC is a binary mixture of felsic and mafic end members as defined by:

$$f + m = 1 \qquad \text{(Eq.1)}$$

$$f \times v_f + m \times v_m = v_{crust} \qquad \text{(Eq. 2)}$$

where $f$ and $m$ are the mass fractions of felsic and mafic end members in the middle (or lower) CC; $v_f$, $v_m$ and $v_{crust}$ are Vp or Vs of the felsic and mafic end members (pressure- and temperature-corrected) and in the crustal layer, respectively. We use only Vp to constrain the felsic fraction ($f$) in the middle or lower CC for three main reasons: using Vs gives results for ($f$) in the deep crust that are in good agreement with those derived from the Vp data, the larger overlap of Vs distributions for the felsic and mafic end-members in the deep crust (Fig. 4b) limits its usefulness in distinguishing the two end-members, and Vs data in the crust are deduced directly from measured Vp data in CRUST 2.0.

Intermediate composition meta-igneous rocks have intermediate seismic velocities compared to those of felsic and mafic rocks, therefore, they are not considered as a separate entity here. As pointed out by *Rudnick and Fountain* [1995], the very large range in velocities for metapelitic sedimentary rocks (metapelites) makes determination of their deep crustal abundances using seismic velocities impossible. Here, we assume that metapelites are a negligible component in the deep crust. Since they have higher abundances of HPEs than mafic rocks and similar HPE contents to felsic rocks, ignoring their presence may lead to an underestimation of HPEs in the deep continental crust. Thus, our estimates should be regarded as minima.

For room temperature and 600 MPa pressure, amphibolite-facies felsic rocks have an average Vp of 6.34±0.16 km/s (1-sigma) and a Vs of 3.65±0.12 km/s, while average mafic amphibolites have a Vp of 6.98±0.20 km/s and a Vs of 3.93±0.15 km/s. Granulite-facies felsic rocks have average Vp of 6.52±0.19 km/s and Vs of 3.70±0.11 km/s, while mafic granulites have average Vp of 7.21±0.20 km/s and Vs of 3.96±0.14 km/s. Our new compilation yields average velocities that are consistent with previous estimates for similar rock types considered by *Christensen and Mooney* [1995] and *Rudnick and Fountain* [1995], but provides a larger sample size than the latter study, due to more recently published laboratory investigations. The sample size considered here is not as large as that reported by *Christensen and Mooney* [1995], who incorporated many unpublished results that are not available to this study.

### 4.3 Rock Type to Chemistry

New and updated compositional databases for amphibolite and granulite facies crustal rocks and mantle peridotites are used here (Appendices B (DOI: 10.1594/IEDA/100245), C (DOI: 10.1594/IEDA/100246) and D) to derive a sample-driven estimate of the average composition of different regions of the continental lithosphere (e.g., amphibolite facies for middle CC, granulite facies for lower CC and xenolithic peridotites for CLM). As with the ultrasonic data compilation, several selection criteria were also applied to the geochemical data compilation. Only data for whole rock samples that were accompanied by appropriate lithological descriptions were used, so that the metamorphic facies of the sample could be properly assigned. X-ray fluorescence determinations of U and Th were excluded due to generally poor data quality, and samples described as being weathered were excluded from the compilation. Finally, major element compositions of all rocks were normalized to 100 wt. % anhydrous, and the log-normal averages of HPEs were adopted, following the recommendation of *Ahrens* [1954], with uncertainties for the average compositions representing the 1-sigma limits.

In addition to the above considerations, intrinsic problems associated with amassing such databases, particularly for peridotites, include [*McDonough*, 1990; *Rudnick et al.*, 1998]:

- Overabundance of data from an individual study, region or laboratory
- Under-representation of some sample types because of their intrinsically lower trace element concentrations (e.g., dunites), presenting a significant analytical challenge (lower limit of detection problems)
- Geological processes (e.g., magmatic entrainment) are potentially non-random processes that may bias our overall view of the deeper portion of the lithosphere
- Weathering can significantly affect the abundances of the mobile elements, particularly K and U.

### 4.3.1 Deep Crust Composition

The compositional databases for amphibolite and granulite facies crustal rocks are both subdivided into felsic, intermediate and mafic meta-igneous rocks based on the normalized $SiO_2$ content, and metasedimentary rocks. For each category, the frequency distributions of HPE abundances show ranges that span nearly four orders of magnitude and are strongly positively skewed, rather than Gaussian (Fig. 5; also see data fitting to metasedimentary rocks in Appendices B and C); they generally fit a log-normal distribution [*Ahrens*, 1954]. In order to decrease the influence of rare enriched or depleted samples on the log-normal average chemical composition for each category, we apply a 1.15-sigma filter that removes ~25% of the data that fall beyond these bounds, and then calculate the central values and associated 1-sigma uncertainties of HPE abundances based on the filtered data for each category (see Supplement Material).

The distributions of the HPE abundances in felsic and mafic amphibolite and granulite facies rocks after such filtering are illustrated in Fig. 6, and the results are reported in Table 5, along with associated 1-sigma uncertainties. These values are adopted in the reference model to estimate the HPE abundances in the heterogeneous middle and lower CC, as described in Section 5.

### 4.3.2 Average Composition of Peridotites and Uncertainties

The peridotite database is subdivided into three categories: spinel, garnet and massif peridotites (Appendix D). Spinel and garnet xenolithic peridotites are assumed to represent the major rock types in the CLM, while massif peridotites are assumed to represent lithospheric mantle under oceanic crust. Due to the analytical challenge of measuring low U and Th concentrations in the lithospheric mantle, there are only several tens of reliable measurements available for statistical analyses of garnet and massif peridotites. We apply the same data treatment (1.15-sigma filtering) to the peridotite database, since distributions of HPEs of all the three types of peridotites are positively skewed and fit the log-normal distribution better than normal distribution. The log-normal mean values adopted in the reference model are close to the median values of the database, and provide robust and coherent estimates to the composition of lithospheric mantle [*McDonough*, 1990; *Rudnick et al.*, 1998] (Table 5).

## 5. Methods of Analysis and Propagation of Uncertainties

We calculate the amount and distribution of HPEs in the Earth (Table 3), which determines the radiogenic heat power and geoneutrino signal of this planet, from the physical (density and thickness) and chemical (abundance of HPEs) characteristics of each reservoir in the reference model. For the middle and lower CC, we use Vp and composition of amphibolite and granulite facies rocks to determine the average abundance of HPEs, as discussed in Sections 4.1 and 4.2 following:

$$a = f \times a_f + m \times a_m \qquad \text{(Eq. 3)}$$

where $a_f$ and $a_m$ is the abundance of HPEs in the felsic and mafic end member, respectively; $a$ is the average abundance in the reservoir. Equations 1 and 2 define the mass fractions of felsic and mafic end members ($f$ and $m$) in the MC and LC reservoirs. In the rare circumstance when the calculated average abundance is more (or less) than the felsic (or mafic) end member, we assume that the average abundance should be the same as the felsic (or mafic) end member. The calculated radiogenic heat power is a direct function of the masses of HPEs and their heat production rates: $9.85\times10^{-5}$, $2.63\times10^{-5}$ and $3.33\times10^{-9}$ W/kg for U, Th and K, respectively [*Dye*, 2012].

The distribution of HPEs in these different reservoirs affects the geoneutrino flux on the Earth's surface. Summing the antineutrino flux produced by HPEs in each volume of our terrestrial model, we calculate the unoscillated geoneutrino flux $\Phi^{(unosc.)}$ expected at the 16 selected sites (Table 1). The flux from U and Th arriving at detectors is smaller than that produced, due to neutrino oscillations, $\Phi^{(osc.)}_{U,\ Th} = \langle Pee \rangle\ \Phi^{(unosc.)}_{U,\ Th}$, where $\langle Pee \rangle = 0.55$ is the average survival probability [*Fiorentini et al.*, 2012]. The geoneutrino event rate in a liquid scintillator detector depends on the number of free protons in the detector, the detection efficiency, the cross section of the inverse beta reaction, and the differential flux of antineutrinos from $^{238}U$ and $^{232}Th$ decay arriving at the detector. Taking into account the U and Th distribution in the Earth, the energy distribution of antineutrinos [*Fiorentini et al.*, 2010], the cross section of inverse beta reaction [*Bemporad et al.*, 2002], and the mass-mixing oscillation parameters [*Fogli et al.*, 2011], we compute the geoneutrino event rate from the decay chain of $^{238}U$ and $^{232}Th$ at four selected sites (Table 2). For simplicity, we neglect the finite energy resolution of the detector, and assume 100% detection efficiency. The expected signal is expressed in TNU (Terrestrial Neutrino Unit), which corresponds to one event per $10^{32}$ target nuclei per year. This unit is commonly used since one kiloton of liquid scintillator contains about $10^{32}$ free protons, and data accumulation takes on the order of several years.

Estimating the uncertainties in the reference model is not straightforward. The commonly used quadratic error propagation method [*Bevington and Robinson*, 2003] is only applicable for linear combinations (addition and subtraction) of errors of normally distributed variables. For non-linear combinations (such as multiplication and division) of uncertainties, the equation provides an approximation when dealing with small uncertainties, and it is derived from the first-order Taylor series expansion applied to the output. Moreover, the error propagation equation cannot be applied when combining asymmetrical uncertainties (non-normal distributions). Because of this, the most common procedure for combining asymmetrical uncertainties is separately tracking the negative error and the positive error using the error propagation equation. This method has no statistical justification and may yield the wrong approximation.

To trace the error propagation in our reference model, we used MATLAB to perform a Monte Carlo simulation [*Huang et al.*, 2013; *Robert and Casella*, 2004; *Rubinstein and Kroese*, 2008]. Monte Carlo simulation is suitable for detailed assessment of uncertainty, particularly when dealing with larger uncertainties, non-normal distributions, and/or complex algorithms. The only requirement for performing Monte Carlo simulation is that the probability functions of all input variables (for example, the abundance of HPEs, seismic velocity, thickness of each layer in the reference model) are determined either from statistical analysis or empirical assumption (see also Supplement Material). Monte Carlo analysis can be performed for any possible shape probability functions, as well as varying degrees of correlation. The Monte Carlo approach consists of three clearly defined steps. The first step is generating large matrices (i.e., $10^4$ random numbers) with pseudorandom samples of input variables according to the specified individual probability functions. Then the matrix of output variable (such as mass of HPEs, radiogenic heat power and geoneutrino flux) with equal size is calculated from the matrixes that are generated following the specified algorithms. The final step is to do statistical analysis of the calculated matrix for the output variable (evaluation of the distribution, central value and uncertainty). The robustness of our results is evaluated by performing iterations to monitor the variation of the output's distribution [*Huang et al.*, 2013]. The relative variations of the central value and 1-sigma uncertainty for the results in this study after performing 100 repeat run with $10^4$ random numbers are about 0.2% and 2%, respectively.

## 6. Discussion

### 6.1 Physical and Chemical Structure of the Reference Crustal Model

The thickness of our reference crustal model is obtained by averaging the three geophysical global crustal models obtained from different approaches, as described in Section 2.2. The distributions of crustal thickness and associated relative uncertainty in our model are shown in Figs. 7a and 7b, respectively. The uncertainties of the continental crustal thickness are not homogeneous: platforms, Archean and Proterozoic shields, the main crustal types composing the interior of stable continents and covering ~50% surface area of the whole CC, have thickness uncertainties ~10%, while the thickness of continental margin crust is more elusive. Larger uncertainties for the thickness estimates occur in the OC, especially for the mid ocean ridges (Fig. 7b). The average crustal thicknesses (including the bulk CC, bulk OC and different continental crustal types) and masses of our reference model are compared with the three geophysical models in Table 6. The global average thicknesses of platforms, and Archean and Proterozoic shields were previously estimated to be 40-43 km [*Christensen and Mooney*, 1995; *Rudnick and Fountain*, 1995]. Although GEMMA yields the thinnest thicknesses for shields and platforms, considering that the typical uncertainty in estimating global average CC thickness is more than 10% [*Čadek and Martinec*, 1991], our reference model, as well as the other three crustal input models, are within uncertainty and equal to that estimated by *Christensen and Mooney* [1995] at the 1-sigma level. Extended crust and orogens in the three input models and in the reference model show average thicknesses higher than, but within 1-sigma of the estimations made by *Christensen and Mooney* [1995]. The surface area weighted average thicknesses of bulk CC for the three input models and our reference model are smaller than the ~41 km estimated by *Christensen and Mooney* [1995], which is likely due to the fact that they did not include continental margins, submerged continental platforms and other thinner crustal types in their compilation. The thickness of OC is generally about 7-8 km, with the exception of the GEMMA model, which yields 8.8 km thick average OC. The possible reason for the thick OC in the GEMMA model is due to the poorly global density distribution under the oceans. However, considering that the average uncertainty in determining the crustal thickness in the oceans is about 2-3 km [*Čadek and Martinec*, 1991], the three input models yield comparable results.

As shown in Fig. 8, the middle and lower CC of our reference model are compositionally heterogeneous on a global scale. The average middle CC derived here has $0.97^{+0.58}_{-0.36}$ μg/g U, $4.86^{+4.30}_{-2.25}$ μg/g Th and $1.52^{+0.81}_{-0.52}$ wt.% K, while the average abundances of U, Th and K in the lower CC are $0.16^{+0.14}_{-0.07}$ μg/g, $0.96^{+1.18}_{-0.51}$ μg/g and $0.65^{+0.34}_{-0.22}$ wt.%, respectively (Table 7; Fig. 8). The uncertainties reported for our new estimates of the HPE abundances in the deep crust are significantly larger than reported in previous global crustal geochemical models [e.g., *Rudnick and Fountain*, 1995; *Rudnick and Gao*, 2003], due to the large dispersions of HPE abundances in amphibolite and granulite facies rocks.

Because of these large uncertainties, all of the estimates for HPEs in the crust of our reference model agree with most previous studies at the 1-sigma level (Table 7). For the middle CC, the central values of our estimates for HPEs are generally only ~10% to 30% lower than those made by *Rudnick and Fountain* [1995] and *Rudnick and Gao* [2003]. For the lower CC, the difference in HPEs between our model and several previous studies is significantly larger than that of the middle CC. Our reference model has lower U and Th, but higher K concentrations, agreeing at the 1-sigma level, than the previous estimates of the lower CC by *Rudnick and Fountain* [1995] and *Rudnick and Gao* [2003]. *Taylor and McLennan* [1995], *McLennan* [2001], *Wedepohl* [1995] and *Hacker et al.* [2011] constructed two-layer crustal models with the top layer being average upper CC (from either their own studies, or *Rudnick and Gao* [2003] in the case of *Hacker et al.* [2011]) and the bottom layer equal to the average of the middle and lower CC in our reference model (see Fig. 3 in *Hacker et al.* [2011]). The abundances of U, Th and K in the combined middle and lower CC of our model are $0.58^{+0.32}_{-0.20}$ μg/g, $2.99^{+2.35}_{-1.35}$ μg/g, and $1.10^{+0.45}_{-0.32}$ wt.%, respectively, which agrees within 1-sigma uncertainty of the estimates of *Hacker et al.* [2011] and within uncertainty of the Th and U abundances of *McLennan* [2001] and of K abundance of *Wedepohl* [1995], but are significantly higher than the abundance estimates of *Taylor and McLennan* [1995] for all elements, and the K abundance estimate of *McLennan* [2001], while lower than the Th and U abundances of [*Wedepohl*, 1995].

In order to compare our estimates of HPE abundance in the bulk CC with previous studies, we recalculate the bulk CC compositions of the other models with the same geophysical crustal structure in our reference model (Table 8). Our estimates of HPE abundances in the bulk CC are close to those determined by *Rudnick and Fountain* [1995], and *Rudnick and Gao* [2003]. Our results also agree with those of *Hacker et al.* [2011], though their Th concentration is at the 1-sigma upper bound of our model. By contrast, our reference model has higher concentrations of K and Th, beyond the 1-sigma level, than estimates by *Taylor and McLennan* [1995], higher K abundance than estimate by *McLennan* [2001], and lower Th abundance than estimate by *Wedepohl* [1995]; the estimates of others are comparable. The fractional masses of U, Th and K concentrated in the bulk CC of our reference model are about 33%, 36% and 28%, respectively, of their total amount in the BSE [*McDonough and Sun*, 1995]. Our estimates of the K/U ($11,621^{+3,512}_{-2,516}$) and Th/U ($4.3^{+1.6}_{-1.0}$) in the bulk CC agree with all previous studies at the 1-sigma level, due to the large uncertainties associated these two ratios derived from large uncertainties of HPE abundance in the CC.

**6.2 Geoneutrino Flux and Radiogenic Heat Power**

In the past decade different authors have presented models for geoneutrino production from the crust, and associated uncertainties. *Mantovani et al.* [2004] adopted minimal and maximal HPE abundances in the literature for each crustal layer of CRUST 2.0 in order to obtain a range of acceptable geoneutrino fluxes. Based on the same CRUST 2.0 model, *Fogli et al.* [2006] and *Dye* [2010] estimated the uncertainties of fluxes based on uncertainties of the HPE abundances reported by *Rudnick and Gao* [2003].

Fig. 9 shows the map of geoneutrino signal at Earth's surface and Fig. 10 illustrates the relative contributions from the convecting mantle (DM+EM) and lithosphere (crust+CLM) to the total surface geoneutrino signals at 16 geographic locations listed in Table 1. In our reference model we estimate the 1-sigma uncertainties of geoneutrino fluxes and radiogenic heat power taking into account two main sources of uncertainties: the physical structure (geophysical uncertainty) and the abundances of HPEs in the reservoirs (geochemical uncertainty). This approach allows us to evaluate the geophysical and geochemical contributions to the uncertainties of our model. With respect to the previous estimates we increase the quality of the predicted geoneutrino signals, pointing out the asymmetrical distributions of the uncertainties as a consequence of the non-Gaussian distributions of HPE abundances in the deep CC and CLM. Within 1-sigma uncertainties, our results for U and Th geoneutrino signals from the crust (Table 2) are comparable to those reported by *Mantovani et al.* [2004] and *Dye* [2010], for which symmetrical and homogeneous uncertainties were adopted. For several locations in Table 2 we report 1-sigma uncertainties of the geoneutrino signal: different relative uncertainties are a consequence of the detailed characterization of the crustal structure and its radioactivity content. From the perspective of deep-Earth exploration based on detection of geoneutrinos from many detectors, our predictions for the lithosphere provide constraints on the signal from the mantle.

The total crustal geoneutrino signal at KamLAND, Borexino and SNO+ are estimated to be $20.6^{+4.0}_{-3.5}$ TNU, $29.0^{+6.0}_{-5.0}$ TNU and $34.0^{+6.3}_{-5.7}$ TNU, respectively, in the reference model. The contributions to the quoted 1-sigma uncertainties from geophysical and geochemical uncertainties can be assessed. By holding the HPE abundances in all crustal reservoirs constant at their central values, the uncertainties associated with the geophysical model are ±1.5 TNU, ±2.7 TNU and ±2.1 TNU, respectively. By fixing the crustal thickness of all voxels as being constant, the geochemical uncertainties contribute $^{+3.6}_{-3.2}$ TNU, $^{+5.0}_{-4.3}$ TNU and $^{+5.9}_{-5.2}$ TNU, respectively. Thus, the geochemical uncertainties clearly dominate the total uncertainty of the crustal geoneutrino signals at all of the three detectors.

Geoneutrino experiments carried on at three existing detectors allow estimation of the geoneutrino flux from the mantle, which, in turn, provides constraints on permissible BSE compositional models [*Dye*, 2010; *Fiorentini et al.*, 2012; *Šrámek et al.*, 2013]. In particular, by subtracting the predicted crustal signal ($S_{crust}$) from the total measured signal ($S_{tot, meas}$) at the three detectors, we can infer the mantle contributions ($S_{mantle}$) for each location [*Fiorentini et al.*, 2012]. These three independently determined mantle signals can be combined to critically evaluate the radiogenic power of the mantle. Furthermore, detailed models of the crustal structure and composition in the region close to the detector show that the uncertainty of the signal from LOcal Crust ($S_{LOC}$, which is dominantly contributed by the 24 1°×1° voxels surrounding the detector) can be reduced when compared to that of a global crustal signal [*Coltorti et al.*, 2011; *Enomoto et al.*, 2007; *Fiorentini et al.*, 2005]. Since $S_{crust}$ in this study is the sum of $S_{LOC}$ and $S_{FFC}$ (the signal from Far Field Crust after excluding local crust), we report in Table 2 the geoneutrino signal $S_{FFC}$ (expected from the Far Field Crust) on the base of our reference model. Thus, at the three existing detectors, one can subtract the $S_{FFC}$ and $S_{LOC}$ from the experimentally measured signal ($S_{tot, meas}$) to define the mantle geoneutrino signals:

$$S_{mantle} = S_{tot, meas} - S_{FFC} - S_{LOC} \qquad (4).$$

The CC in the reference model contributes $6.8^{+1.4}_{-1.1}$ TW radiogenic heat power to the total 20.1 TW radiogenic power generated in the BSE, which agrees with previous estimates by *Hacker et al.* [2011], *McLennan* [2001], *Rudnick and Fountain* [1995], and *Rudnick and Gao* [2003] at the 1-sigma level, but is higher than estimate by *Taylor and McLennan* [1995], and lower than estimate by *Wedepohl* [1995] (Table 7). We estimate a 1-sigma uncertainty of ±0.8 TW and $^{+1.1}_{-0.8}$ TW of the radiogenic heat power of the CC corresponding to geophysical and geochemical uncertainty in our reference model, respectively.

Although the mass of OC (excluding the overlying sediment) is poorly known, its contribution to the anticipated geoneutrino signals at the three existing detectors is less than 0.2 TNU at the 1-sgima level. By contrast, we calculate that the CLM contributes 1.6 TNU, 2.2 TNU and 2.1 TNU to the geoneutrino signal at KamLAND, Borexino and SNO+, respectively (Table 2).

The uncertainties associated with the signal coming from this portion of lithosphere are large, and an increase in signal by a factor three is permitted at the 1-sigma level. Determining the distribution of U and Th in the lithospheric mantle sections underlying the detectors would thus be desirable in the future. Despite the fact that the mass of the CLM is about five times the crustal mass, it contains approximately 10% of the total mass of HPEs in the crust. The radiogenic heat power of CLM is $0.8^{+1.1}_{-0.6}$ TW: the main contribution to the uncertainty comes from the large 1-sigma uncertainty of HPE abundances in peridotites.

**7. Conclusions**

In this paper we provide a reference model for the geoneutrino flux and radiogenic heat power from the main reservoirs of our planet. A particular effort has been dedicated to estimating uncertainties derived from the geophysical constrains and from the geochemical data. We summarize here the main results reached in this study.

1. Three geophysical global crustal models based on reflection and refraction seismic body wave (CRUST 2.0), surface wave dispersion (CUB 2.0), and gravimetric anomalies (GEMMA) are studied with the aim to estimate the geophysical uncertainties of our reference crustal model. It yields an average crustal thickness of 34.4±4.1 km in the continents and 8.0±2.7 km in the oceans. Moreover a global map of the uncertainties associated to the crustal thickness has been produced with a grid of 1°×1° voxel.

2. The average continental crust derived here contains $1.31^{+0.29}_{-0.25}$ μg/g U, $5.61^{+1.56}_{-0.89}$ μg/g Th and $1.52^{+0.29}_{-0.22}$ wt. % K, has $\mathrm{Th/U} = 4.3^{+1.6}_{-1.0}$, $\mathrm{K/U} = 11,621^{+3,512}_{-2,516}$ and produces $6.8^{+1.4}_{-1.1}$ TW of heat. These asymmetrical uncertainties are propagated from the non-Gaussian distributions of HPE abundances in the deep continental crust and continental lithospheric mantle using Monte Carlo simulation.

3. The radiogenic heat power in different Earth reservoirs and the geoneutrino flux at 16 geographic locations are calculated with consideration of two main sources of uncertainties: the physical structure (geophysical uncertainty) and the abundances of HPEs in the reservoirs (geochemical uncertainty). Contributions from the two different sources of uncertainty to the global uncertainties are estimated for the first time, and we show that the geochemical uncertainty exerts the greatest control on the overall uncertainties.

4. The geoneutrino flux from the continental lithospheric mantle (CLM) is calculated here for the first-time based on an updated xenolithic peridotite database. The calculated geoneutrino signal from the CLM exceeds that from the oceanic crust (OC) at all three existing detectors.

5. The combination of this global crust model, detailed local crust models, and the measured signal for each detector, provide the critical inputs needed to assess the global mantle signal and its uncertainty. Thus, the mantle signal at each detector and its uncertainty can be independently combined to place limits on acceptable models for the mantle's radiogenic power.

**Acknowledgements**

We thank Vedran Lekic for providing the dataset of CUB 2.0 and for insightful discussions about the seismic models. We are grateful to Mirko Reguzzoni and Daniele Sampietro for sharing the GEMMA data. Discussions with Mohammad Bagherbandi about Moho Density Contrast model [*Sjöberg and Bagherbandi*, 2011] helped us to understand the gravitational anomaly model. Andrew Kerr kindly provided a geochemical database for basaltic oceanic plateaus. We appreciate statistical insights from Michael Evans, Eligio Lisi and Gerti Xhixha. Thoughtful comments from William M. White and an anonymous reviewer are sincerely appreciated. This study was funded by National Science Foundation Grant EAR-1067983/1068097 and partially by National Institute of Nuclear Physics (INFN).

Table 1: Geoneutrino flux (no oscillation) from U, Th, K in the lithosphere (LS; crust+CLM), upper depleted mantle (DM) and lower enriched mantle (EM) for 16 geographic sites. The unit of flux is $10^6$ $cm^{-2}\cdot s^{-1}$. Uncertainties are 1 sigma.

| Site | Kamioka, JP* 36.43 N, 137.31 E[a] | | | Gran Sasso, IT 42.45 N, 13.57 E[b] | | | Sudbury, CA 46.47 N, 81.20 W[c] | | | Hawaii, US 19.72 N, 156.32 W[d] | | |
|---|---|---|---|---|---|---|---|---|---|---|---|---|
| | Φ(U) | Φ(Th) | Φ(K) | Φ(U) | Φ(Th) | Φ(K) | Φ(U) | Φ(Th) | Φ(K) | Φ(U) | Φ(Th) | Φ(K) |
| LS | $2.47^{+0.65}_{-0.53}$ | $2.25^{+0.67}_{-0.43}$ | $9.42^{+2.19}_{-1.65}$ | $3.36^{+0.96}_{-0.75}$ | $3.46^{+1.26}_{-0.80}$ | $14.83^{+3.99}_{-2.96}$ | $3.92^{+0.98}_{-0.78}$ | $3.78^{+1.17}_{-0.73}$ | $16.16^{+3.67}_{-2.76}$ | $0.32^{+0.10}_{-0.08}$ | $0.31^{+0.11}_{-0.07}$ | $1.30^{+0.36}_{-0.28}$ |
| DM | 0.59 | 0.36 | 4.08 | 0.57 | 0.35 | 3.96 | 0.58 | 0.35 | 3.96 | 0.63 | 0.38 | 4.34 |
| EM | 0.41 | 0.42 | 1.75 | 0.41 | 0.42 | 1.75 | 0.41 | 0.42 | 1.75 | 0.41 | 0.42 | 1.75 |
| Total | $3.47^{+0.65}_{-0.53}$ | $3.03^{+0.67}_{-0.43}$ | $15.25^{+2.19}_{-1.65}$ | $4.34^{+0.96}_{-0.75}$ | $4.23^{+1.26}_{-0.80}$ | $20.54^{+3.99}_{-2.96}$ | $4.90^{+0.98}_{-0.78}$ | $4.55^{+1.17}_{-0.73}$ | $21.88^{+3.67}_{-2.76}$ | $1.36^{+0.10}_{-0.08}$ | $1.11^{+0.11}_{-0.07}$ | $7.40^{+0.36}_{-0.28}$ |
| Site | Baksan, RU 43.20 N, 42.72 E[e] | | | Homestake, US 44.35 N, 103.75 W[f] | | | Pyhasalmi, FI 63.66 N, 26.05 E[g] | | | Curacao, NA 12.00 N, 69.00 W[h] | | |
| | Φ(U) | Φ(Th) | Φ(K) | Φ(U) | Φ(Th) | Φ(K) | Φ(U) | Φ(Th) | Φ(K) | Φ(U) | Φ(Th) | Φ(K) |
| LS | $4.13^{+1.11}_{-0.90}$ | $3.95^{+1.26}_{-0.83}$ | $16.14^{+4.02}_{-3.00}$ | $4.28^{+1.17}_{-0.97}$ | $4.13^{+1.34}_{-0.87}$ | $16.96^{+4.37}_{-3.13}$ | $4.00^{+1.00}_{-0.83}$ | $3.64^{+1.02}_{-0.67}$ | $15.35^{+3.33}_{-2.50}$ | $2.21^{+0.63}_{-0.47}$ | $2.10^{+0.68}_{-0.44}$ | $8.69^{+2.19}_{-1.64}$ |
| DM | 0.57 | 0.34 | 3.92 | 0.58 | 0.35 | 3.96 | 0.57 | 0.35 | 3.93 | 0.59 | 0.36 | 4.06 |
| EM | 0.41 | 0.42 | 1.75 | 0.41 | 0.42 | 1.75 | 0.41 | 0.42 | 1.75 | 0.41 | 0.42 | 1.75 |
| Total | $5.10^{+1.11}_{-0.90}$ | $4.36^{+1.26}_{-0.83}$ | $21.81^{+4.02}_{-3.00}$ | $5.26^{+1.17}_{-0.97}$ | $4.90^{+1.34}_{-0.87}$ | $22.68^{+4.37}_{-3.13}$ | $4.98^{+1.00}_{-0.83}$ | $4.41^{+1.02}_{-0.67}$ | $21.04^{+3.33}_{-2.50}$ | $3.20^{+0.63}_{-0.47}$ | $2.88^{+0.68}_{-0.44}$ | $14.51^{+2.19}_{-1.64}$ |
| Site | Canfranc, SP 42.70 N, 0.52 W[i] | | | Fréjus, FR 45.13 N, 6.68 E[i] | | | Slanic, RO 45.23 N, 25.94 E[i] | | | SUNLAB, PL 51.55 N, 16.03 E[i] | | |
| | Φ(U) | Φ(Th) | Φ(K) | Φ(U) | Φ(Th) | Φ(K) | Φ(U) | Φ(Th) | Φ(K) | Φ(U) | Φ(Th) | Φ(K) |
| LS | $3.36^{+0.90}_{-0.76}$ | $3.27^{+1.07}_{-0.71}$ | $13.70^{+3.50}_{-2.57}$ | $3.58^{+1.12}_{-0.89}$ | $3.62^{+1.39}_{-0.89}$ | $15.27^{+4.34}_{-3.37}$ | $3.87^{+1.10}_{-0.91}$ | $3.81^{+1.30}_{-0.85}$ | $16.13^{+4.32}_{-3.21}$ | $3.67^{+0.96}_{-0.78}$ | $3.69^{+1.29}_{-0.80}$ | $15.88^{+3.94}_{-3.00}$ |
| DM | 0.58 | 0.35 | 3.97 | 0.58 | 0.35 | 3.96 | 0.57 | 0.34 | 3.93 | 0.57 | 0.35 | 3.93 |
| EM | 0.41 | 0.42 | 1.75 | 0.41 | 0.42 | 1.75 | 0.41 | 0.42 | 1.75 | 0.41 | 0.42 | 1.75 |
| Total | $4.34^{+0.90}_{-0.76}$ | $4.04^{+1.07}_{-0.71}$ | $19.43^{+3.50}_{-2.57}$ | $4.56^{+1.12}_{-0.89}$ | $4.39^{+1.39}_{-0.89}$ | $20.99^{+4.34}_{-3.37}$ | $4.85^{+1.10}_{-0.91}$ | $4.58^{+1.30}_{-0.85}$ | $21.81^{+4.32}_{-3.21}$ | $4.65^{+0.96}_{-0.78}$ | $4.46^{+1.29}_{-0.80}$ | $21.57^{+3.94}_{-3.00}$ |
| Site | Boulby, UK 54.55 N, 0.82 W[i] | | | Daya Bay, CH 23.13 N, 114.67 E[j] | | | Himalaya, CH 33.00 N, 85.00 E | | | Rurutu, FP 22.47 S, 151.33 W | | |
| | Φ(U) | Φ(Th) | Φ(K) | Φ(U) | Φ(Th) | Φ(K) | Φ(U) | Φ(Th) | Φ(K) | Φ(U) | Φ(Th) | Φ(K) |
| LS | $3.24^{+0.84}_{-0.68}$ | $3.16^{+1.03}_{-0.65}$ | $13.44^{+3.19}_{-2.40}$ | $3.06^{+0.85}_{-0.70}$ | $2.95^{+1.02}_{-0.63}$ | $12.69^{+3.11}_{-2.40}$ | $5.63^{+1.61}_{-1.34}$ | $5.48^{+1.88}_{-1.22}$ | $23.05^{+6.24}_{-4.69}$ | $0.27^{+0.08}_{-0.07}$ | $0.25^{+0.09}_{-0.06}$ | $1.09^{+0.30}_{-0.24}$ |
| DM | 0.57 | 0.35 | 3.96 | 0.58 | 0.35 | 3.99 | 0.57 | 0.35 | 3.93 | 0.63 | 0.38 | 4.36 |
| EM | 0.41 | 0.42 | 1.75 | 0.41 | 0.42 | 1.75 | 0.41 | 0.42 | 1.75 | 0.41 | 0.42 | 1.75 |
| Total | $4.22^{+0.84}_{-0.68}$ | $3.93^{+1.03}_{-0.65}$ | $19.15^{+3.19}_{-2.40}$ | $4.04^{+0.85}_{-0.70}$ | $3.72^{+1.02}_{-0.63}$ | $18.43^{+3.11}_{-2.40}$ | $6.61^{+1.61}_{-1.34}$ | $6.25^{+1.88}_{-1.22}$ | $28.73^{+6.24}_{-4.69}$ | $1.31^{+0.08}_{-0.07}$ | $1.05^{+0.09}_{-0.06}$ | $7.20^{+0.30}_{-0.24}$ |

*JP: Japan; IT: Italy; CA: Canada; US: United States of America; RU: Russia; FI: Finland; NA: Netherlands Antilles; SP: Spain; FR: France; RO: Romania; PL: Poland; UK: United Kingdom; CH: China; FP: French Polynesia

[a]*Araki et al.*, 2005; [b]*Alvarez Sanchez et al.*, 2012; [c]*Chen*, 2006; [d]*Dye*, 2010;[e]*Buklerskii et al.*, 1995; [f]*Tolich et al.*, 2006; [g]*Wurm et al.*, 20012; [h]*De Meijer et al.*, 2006; [i]LAGUNA project website; [j]*Wang*, 2011

Table 2: Geoneutrino signal at four selected sites from each reservoir as described in our reference Earth model. The unit of signal is TNU as defined in Section 4.5.

| | KamLAND 36.43 N, 137.31 E | | | Borexino 42.45 N, 13.57 E | | | SNO+ 46.47 N, 81.20 W | | | Hanohano 19.72 N, 156.32 W | | |
|---|---|---|---|---|---|---|---|---|---|---|---|---|
| | S(U) | S(Th) | S(U+Th)[a] | S(U) | S(Th) | S(U+Th) | S(U) | S(Th) | S(U+Th) | S(U) | S(Th) | S(U+Th) |
| Sed_CC | $0.3^{+0.1}_{-0.1}$ | $0.10^{+0.02}_{-0.02}$ | $0.4^{+0.1}_{-0.1}$ | $1.3^{+0.3}_{-0.3}$ | $0.4^{+0.1}_{-0.1}$ | $1.8^{+0.3}_{-0.3}$ | $0.4^{+0.1}_{-0.1}$ | $0.12^{+0.02}_{-0.02}$ | $0.5^{+0.1}_{-0.1}$ | $0.08^{+0.02}_{-0.01}$ | $0.03^{+0.01}_{-0.01}$ | $0.12^{+0.02}_{-0.02}$ |
| UC | $11.7^{+3.0}_{-2.7}$ | $3.2^{+0.4}_{-0.4}$ | $15.0^{+3.0}_{-2.7}$ | $13.7^{+3.6}_{-3.4}$ | $3.7^{+0.6}_{-0.5}$ | $17.5^{+3.6}_{-3.4}$ | $18.2^{+4.4}_{-4.3}$ | $4.9^{+0.6}_{-0.6}$ | $23.3^{+4.4}_{-4.3}$ | $1.1^{+0.3}_{-0.3}$ | $0.3^{+0.05}_{-0.04}$ | $1.4^{+0.3}_{-0.3}$ |
| MC | $2.9^{+2.0}_{-1.2}$ | $0.9^{+1.0}_{-0.5}$ | $4.0^{+2.4}_{-1.5}$ | $4.7^{+3.3}_{-1.8}$ | $1.8^{+1.7}_{-0.9}$ | $7.0^{+3.9}_{-2.5}$ | $5.5^{+3.6}_{-2.1}$ | $1.9^{+1.9}_{-0.9}$ | $7.9^{+4.2}_{-2.7}$ | $0.4^{+0.3}_{-0.2}$ | $0.1^{+0.1}_{-0.1}$ | $0.6^{+0.3}_{-0.2}$ |
| LC | $0.4^{+0.5}_{-0.3}$ | $0.1^{+0.2}_{-0.1}$ | $0.5^{+0.5}_{-0.3}$ | $1.0^{+0.8}_{-0.4}$ | $0.5^{+0.9}_{-0.3}$ | $1.7^{+1.3}_{-0.7}$ | $0.9^{+0.7}_{-0.4}$ | $0.4^{+0.6}_{-0.2}$ | $1.4^{+1.1}_{-0.6}$ | $0.06^{+0.05}_{-0.03}$ | $0.03^{+0.03}_{-0.01}$ | $0.1^{+0.07}_{-0.04}$ |
| CLM | $1.0^{+1.9}_{-0.7}$ | $0.3^{+0.7}_{-0.2}$ | $1.6^{+2.2}_{-1.0}$ | $1.4^{+2.7}_{-1.0}$ | $0.4^{+1.0}_{-0.3}$ | $2.2^{+3.1}_{-1.3}$ | $1.3^{+2.5}_{-0.9}$ | $0.4^{+0.9}_{-0.3}$ | $2.1^{+2.9}_{-1.2}$ | $0.2^{+0.3}_{-0.1}$ | $0.06^{+0.12}_{-0.04}$ | $0.3^{+0.4}_{-0.2}$ |
| Sed_OC | $0.1^{+0.1}_{-0.1}$ | $0.04^{+0.02}_{-0.02}$ | $0.2^{+0.1}_{-0.1}$ | $0.2^{+0.1}_{-0.1}$ | $0.06^{+0.02}_{-0.02}$ | $0.2^{+0.1}_{-0.1}$ | $0.09^{+0.03}_{-0.03}$ | $0.03^{+0.01}_{-0.01}$ | $0.13^{+0.03}_{-0.03}$ | $0.08^{+0.03}_{-0.03}$ | $0.03^{+0.01}_{-0.01}$ | $0.12^{+0.03}_{-0.03}$ |
| OC | $0.1^{+0.1}_{-0.1}$ | $0.02^{+0.01}_{-0.01}$ | $0.1^{+0.1}_{-0.1}$ | $0.05^{+0.02}_{-0.02}$ | $0.01^{+0.01}_{-0.00}$ | $0.06^{+0.02}_{-0.02}$ | $0.04^{+0.02}_{-0.02}$ | $0.01^{+0.00}_{-0.00}$ | $0.05^{+0.02}_{-0.02}$ | $0.3^{+0.1}_{-0.1}$ | $0.05^{+0.03}_{-0.02}$ | $0.3^{+0.1}_{-0.1}$ |
| **Bulk Crust** | $15.6^{+3.7}_{-3.2}$ | $4.5^{+1.0}_{-0.7}$ | $20.6^{+4.0}_{-3.5}$ | $21.4^{+5.2}_{-4.6}$ | $6.8^{+2.3}_{-1.4}$ | $29.0^{+6.0}_{-5.0}$ | $25.7^{+5.9}_{-5.2}$ | $7.7^{+2.2}_{-1.3}$ | $34.0^{+6.3}_{-5.7}$ | $1.7^{+0.4}_{-0.3}$ | $0.5^{+0.2}_{-0.1}$ | $2.6^{+0.5}_{-0.5}$ |
| **FFC**[b] | $5.5^{+1.4}_{-1.2}$ | $1.7^{+0.5}_{-0.3}$ | $7.3^{+1.5}_{-1.2}$ | $10.3^{+2.6}_{-2.2}$ | $3.2^{+1.1}_{-0.7}$ | $13.7^{+2.8}_{-2.3}$ | $11.5^{+2.7}_{-2.3}$ | $3.4^{+1.0}_{-0.6}$ | $15.1^{+2.8}_{-2.4}$ | -- | -- | -- |
| **Total LS**[c] | $17.5^{+4.6}_{-3.8}$ | $5.0^{+1.5}_{-1.0}$ | $22.7^{+4.9}_{-4.1}$ | $23.6^{+6.8}_{-5.2}$ | $7.6^{+2.9}_{-1.8}$ | $31.9^{+7.3}_{-5.8}$ | $27.8^{+6.9}_{-5.7}$ | $8.4^{+2.7}_{-1.7}$ | $36.7^{+7.5}_{-6.3}$ | $2.3^{+0.7}_{-0.5}$ | $0.7^{+0.2}_{-0.2}$ | $3.0^{+0.7}_{-0.6}$ |
| DM | 4.2 | 0.8 | 5.0 | 4.0 | 0.8 | 4.9 | 4.1 | 0.8 | 4.9 | 4.4 | 0.8 | 5.2 |
| EM | 2.9 | 0.9 | 3.8 | 2.9 | 0.9 | 3.8 | 2.9 | 0.9 | 3.8 | 2.9 | 0.9 | 3.8 |
| **Grand Total** | $31.5^{+4.9}_{-4.1}$ | | -- | $40.3^{+7.3}_{-5.8}$ | | -- | $45.4^{+7.5}_{-6.3}$ | | -- | $12.0^{+0.7}_{-0.6}$ | | -- |

[a]The sum of signals from U and Th is obtained by Monte Carlo simulation; all the reported uncertainties are 1 sigma.
[b]FFC is defined as Far Field Crust with the geoneutrino signal originated from the 24 closest 1°×1° crustal voxels excluded from the bulk crustal signal (see Section 6.2).
[c]LS: lithosphere; defined as CC+OC+CLM.

Table 3: Global average physical (density, thickness, mass and radiogenic heat power) and chemical (abundance and mass of HPEs) properties of each reservoir as described in the reference model.

| | | ρ, $g/cm^3$ | d, km | M, $10^{21}$ kg | Abundance | | | Mass | | | H, TW |
|---|---|---|---|---|---|---|---|---|---|---|---|
| | | | | | U, μg/g | Th, μg/g | K, % | U, $10^{15}$ kg | Th, $10^{15}$ kg | K, $10^{19}$ kg | |
| CC | Sed | 2.25[a] | 1.5±0.3 | 0.7±0.1 | 1.73±0.09 | 8.10±0.59 | 1.83±0.12 | $1.2^{+0.2}_{-0.2}$ | $5.8^{+1.1}_{-1.1}$ | $1.3^{+0.2}_{-0.2}$ | $0.3^{+0.1}_{-0.1}$ |
| | UC | 2.76 | 11.6±1.3 | 6.7±0.8 | 2.7±0.6 | 10.5±1.0 | 2.32±0.19 | $18.2^{+4.8}_{-4.3}$ | $70.7^{+10.7}_{-10.2}$ | $15.6^{+2.3}_{-2.1}$ | $4.2^{+0.7}_{-0.6}$ |
| | MC | 2.88 | 11.4±1.3 | 6.9±0.9 | $0.97^{+0.58}_{-0.36}$ | $4.86^{+4.30}_{-2.25}$ | $1.52^{+0.81}_{-0.52}$ | $6.6^{+4.1}_{-2.5}$ | $33.3^{+30.0}_{-15.5}$ | $10.4^{+5.7}_{-3.7}$ | $1.9^{+0.9}_{-0.6}$ |
| | LC | 3.05 | 10.0±1.2 | 6.3±0.7 | $0.16^{+0.14}_{-0.07}$ | $0.96^{+1.18}_{-0.51}$ | $0.65^{+0.34}_{-0.22}$ | $1.0^{+0.9}_{-0.4}$ | $6.0^{+7.7}_{-3.3}$ | $4.1^{+2.2}_{-1.4}$ | $0.4^{+0.3}_{-0.1}$ |
| | LM | 3.37 | 140±71 | 97±47 | $0.03^{+0.05}_{-0.02}$ | $0.15^{+0.28}_{-0.10}$ | $0.03^{+0.04}_{-0.02}$ | $2.9^{+5.4}_{-2.0}$ | $14.5^{+29.4}_{-9.4}$ | $3.1^{+4.7}_{-1.8}$ | $0.8^{+1.1}_{-0.6}$ |
| OC | Sed | 2.03 | 0.6±0.2 | 0.3±0.1 | 1.73±0.09 | 8.10±0.59 | 1.83±0.12 | $0.6^{+0.2}_{-0.2}$ | $2.8^{+0.9}_{-0.9}$ | $0.6^{+0.2}_{-0.2}$ | $0.2^{+0.1}_{-0.1}$ |
| | C | 2.88 | 7.4±2.6 | 6.3±2.2 | 0.07±0.02 | 0.21±0.06 | 0.07±0.02 | $0.4^{+0.2}_{-0.2}$ | $1.3^{+0.7}_{-0.5}$ | $0.4^{+0.2}_{-0.2}$ | $0.1^{+0.04}_{-0.03}$ |
| DM[b] | | 4.66 | 2090 | 3207 | 0.008 | 0.022 | 0.015 | 25.7 | 70.6 | 48.7 | 6.0 |
| EM[c] | | 5.39 | 710 | 704 | 0.034 | 0.162 | 0.041 | 24.0 | 113.7 | 28.7 | 6.3 |
| BSE[d] | | 4.42 | 2891 | 4035 | 0.020 | 0.079 | 0.028 | 80.7 | 318.8 | 113.0 | 20.1 |

[a]The uncertainty in density is about the same as that of Vp (3-4%) [*Mooney et al.*, 1998]. All other reported uncertainties are 1 sigma.
[b]The physical structure of the mantle is based on PREM; HPE abundances in DM are derived from *Arevalo and McDonough* [2010].
[c]HPE abundances in EM are calculated through a mass balance of HPEs in the mantle, with EM has a mass ~18% of the total mass of the convecting mantle.
[d]BSE composition of *McDonough and Sun* [1995].

Table 4: Average properties of amphibolite and granulite facies rocks for density, $SiO_2$ contents and laboratory-measured Vp and Vs at 600 MPa and room temperature.

| | All Samples | | | Samples for which Vs is available | | | |
|---|---|---|---|---|---|---|---|
| | Density *g/cm³* | $SiO_2$ *wt. %* | Vp *km/s* | Density *g/cm³* | $SiO_2$ *wt. %* | Vp *km/s* | Vs *km/s* |
| | **Amphibolite Facies** | | | | | | |
| **Felsic** | | | | | | | |
| N[a] | 77 | 50 | 77 | 36 | 31 | 36 | 36 |
| mean | 2.719 | 69.19 | **6.34**[b] | 2.751 | 68.91 | 6.30 | 3.65 |
| standard deviation | 0.084 | 3.51 | **0.16** | 0.075 | 3.81 | 0.17 | 0.12 |
| median | 2.703 | 69.98 | 6.30 | 2.737 | 69.83 | 6.26 | 3.66 |
| **Intermediate** | | | | | | | |
| N | 20 | 19 | 20 | 11 | 11 | 11 | 11 |
| mean | 2.856 | 56.65 | 6.62 | 2.857 | 56.83 | 6.56 | 3.72 |
| standard deviation | 0.085 | 3.95 | 0.26 | 0.091 | 4.12 | 0.30 | 0.23 |
| median | 2.850 | 54.80 | 6.67 | 2.854 | 54.80 | 6.48 | 3.73 |
| **Mafic** | | | | | | | |
| N | 57 | 43 | 57 | 34 | 26 | 34 | 34 |
| mean | 3.036 | 48.26 | **6.98** | 3.059 | 48.03 | 6.96 | 3.93 |
| standard deviation | 0.068 | 1.91 | **0.20** | 0.069 | 2.15 | 0.20 | 0.15 |
| median | 3.030 | 48.10 | 6.99 | 3.077 | 47.82 | 6.94 | 3.95 |
| **Metapelite** | | | | | | | |
| N | 27 | 21 | 44 | 7 | 4 | 18 | 18 |
| mean | 2.772 | 64.14 | 6.45 | 2.849 | 58.89 | 6.48 | 3.63 |
| standard deviation | 0.090 | 7.40 | 0.21 | 0.080 | 8.91 | 0.17 | 0.13 |
| median | 2.751 | 65.08 | 6.46 | 2.864 | 62.25 | 6.47 | 3.63 |
| | **Granulite Facies** | | | | | | |
| **Felsic** | | | | | | | |
| N | 29 | 27 | 29 | 12 | 10 | 12 | 12 |
| mean | 2.715 | 68.89 | **6.52** | 2.760 | 67.77 | 6.47 | 3.70 |
| standard deviation | 0.072 | 4.24 | **0.19** | 0.071 | 5.38 | 0.18 | 0.11 |
| median | 2.694 | 68.30 | 6.51 | 2.773 | 65.42 | 6.48 | 3.69 |
| **Intermediate** | | | | | | | |
| N | 12 | 9 | 12 | 10 | 7 | 10 | 10 |
| mean | 2.895 | 56.27 | 6.74 | 2.886 | 56.03 | 6.69 | 3.67 |
| standard deviation | 0.105 | 3.44 | 0.17 | 0.107 | 3.39 | 0.11 | 0.16 |
| median | 2.898 | 54.30 | 6.71 | 2.896 | 54.30 | 6.71 | 3.71 |
| **Mafic** | | | | | | | |
| N | 44 | 40 | 44 | 32 | 28 | 32 | 32 |
| mean | 3.066 | 47.19 | **7.21** | 3.079 | 47.11 | 7.19 | 3.96 |
| standard deviation | 0.112 | 1.98 | **0.20** | 0.122 | 2.12 | 0.23 | 0.14 |
| median | 3.067 | 47.23 | 7.23 | 3.085 | 47.00 | 7.23 | 3.98 |
| **Metapelite** | | | | | | | |
| N | 21 | 16 | 23 | 17 | 12 | 18 | 18 |
| mean | 3.059 | 53.23 | 6.98 | 3.067 | 53.31 | 6.90 | 3.99 |
| standard deviation | 0.137 | 5.29 | 0.43 | 0.150 | 5.89 | 0.43 | 0.18 |
| median | 3.064 | 52.13 | 7.03 | 3.074 | 52.13 | 6.97 | 3.99 |

[a]N is the number of samples compiled in the dataset.
[b]Bold numbers are the Vp of felsic and mafic end members in middle and lower CC used in the reference model.

Table 5: Average HPE abundances in amphibolite facies, granulite facies and peridotite rocks. '+' represents the upper uncertainty and '-' represents the lower uncertainty.

| | **$K_2O$** | *1 sigma* | | | | **Th** | *1 sigma* | | | | **U** | *1 sigma* | | | |
|---|---|---|---|---|---|---|---|---|---|---|---|---|---|---|---|
| | Mean[a] | + | - | Median | n | Mean | + | - | Median | n | Mean | + | - | Median | n |
| | **Amphibolite Facies (MC)** | | | | | | | | | | | | | | |
| Felsic All[b] | 2.41 | 2.83 | 1.30 | 2.97 | 670 | 6.60 | 15.17 | 4.60 | 8.98 | 534 | 1.25 | 2.02 | 0.77 | 1.39 | 485 |
| Felsic 1.15[c] | **2.89**[d] | **1.81** | **1.11** | 3.19 | 578 | **8.27** | **8.12** | **4.10** | 9.43 | 428 | **1.37** | **1.03** | **0.59** | 1.43 | 368 |
| Intermediate All | 0.96 | 1.82 | 0.63 | 1.22 | 324 | 1.90 | 5.53 | 1.41 | 2.50 | 185 | 0.63 | 1.10 | 0.40 | 0.66 | 166 |
| Intermediate 1.15 | 1.15 | 1.09 | 0.56 | 1.28 | 245 | 2.22 | 2.87 | 1.25 | 2.70 | 138 | 0.73 | 0.55 | 0.31 | 0.76 | 128 |
| Metapelitic All | 2.27 | 3.52 | 1.38 | 2.89 | 298 | 6.36 | 13.70 | 4.34 | 8.97 | 224 | 1.68 | 3.13 | 1.09 | 2.00 | 199 |
| Metapelitic 1.15 | 2.84 | 1.54 | 1.00 | 2.96 | 269 | 8.14 | 6.48 | 3.61 | 9.45 | 200 | 1.95 | 1.28 | 0.77 | 2.07 | 173 |
| Mafic All | 0.48 | 0.79 | 0.30 | 0.52 | 569 | 0.62 | 1.29 | 0.42 | 0.60 | 340 | 0.34 | 0.69 | 0.23 | 0.37 | 303 |
| Mafic 1.15 | **0.50** | **0.41** | **0.23** | 0.53 | 420 | **0.58** | **0.57** | **0.29** | 0.57 | 257 | **0.37** | **0.39** | **0.19** | 0.39 | 233 |
| | **Granulite Facies (LC)** | | | | | | | | | | | | | | |
| Felsic All | 2.19 | 3.06 | 1.28 | 2.66 | 719 | 3.03 | 13.38 | 2.47 | 4.08 | 177 | 0.40 | 0.83 | 0.27 | 0.48 | 141 |
| Felsic 1.15 | **2.71** | **2.05** | **1.17** | 3.15 | 568 | **3.87** | **7.35** | **2.54** | 4.80 | 133 | **0.42** | **0.41** | **0.21** | 0.48 | 108 |
| Intermediate All | 0.95 | 1.33 | 0.56 | 0.94 | 535 | 0.49 | 2.46 | 0.41 | 0.31 | 208 | 0.12 | 0.36 | 0.09 | 0.10 | 173 |
| Intermediate 1.15 | 0.95 | 0.60 | 0.37 | 0.91 | 383 | 0.36 | 0.77 | 0.25 | 0.29 | 166 | 0.10 | 0.12 | 0.05 | 0.10 | 130 |
| Metapelitic All | 1.61 | 2.71 | 1.01 | 2.22 | 294 | 3.04 | 15.66 | 2.55 | 6.30 | 119 | 0.56 | 0.93 | 0.35 | 0.60 | 89 |
| Metapelitic 1.15 | 2.11 | 1.54 | 0.89 | 2.42 | 247 | 5.44 | 11.60 | 3.70 | 7.90 | 91 | 0.59 | 0.41 | 0.24 | 0.60 | 69 |
| Mafic All | 0.36 | 0.63 | 0.23 | 0.40 | 780 | 0.33 | 1.22 | 0.26 | 0.32 | 328 | 0.11 | 0.36 | 0.08 | 0.12 | 286 |
| Mafic 1.15 | **0.39** | **0.31** | **0.17** | 0.40 | 579 | **0.30** | **0.46** | **0.18** | 0.30 | 258 | **0.10** | **0.14** | **0.06** | 0.11 | 236 |
| | **Peridotite (LM)** | | | | | | | | | | | | | | |
| Peridotite All | 0.044 | 0.112 | 0.031 | 0.040 | 916 | 0.122 | 0.689 | 0.104 | 0.150 | 233 | 0.027 | 0.113 | 0.022 | 0.033 | 149 |
| Peridotite 1.15 | **0.038** | **0.052** | **0.022** | 0.040 | 752 | **0.150** | **0.277** | **0.097** | 0.165 | 184 | **0.033** | **0.049** | **0.020** | 0.028 | 118 |

[a]log-normal mean, $K_2O$ concentration is in wt.%, Th and U concentrations are in µg/g. [b]"All" results are from all compiled data.
[c]"1.15" results are from filtered data within 1.15 sigma of the log-normal distribution.
[d]Bold numbers are used for determining the amount and distribution of HPEs in the middle and lower CC and CLM.

Table 6: Comparison of crustal thickness and mass between the three global crustal models (CRUST 2.0, CUB 2.0 and GEMMA) and our reference model (**RM**).

| | | Area[a](%) | CRUST 2.0 | CUB 2.0 | GEMMA | **RM**[b] | CM’95[c] |
|---|---|---|---|---|---|---|---|
| Thickness (km) | Platform | 14 | 41.0 | 40.4 | 36.3 | 39.2±4.2 | 41.5 |
| | Archean Shield | 20 | 37.9 | 38.1 | 36.6 | 37.5±3.1 | 41.5 |
| | Proterozoic Shield | 15 | 40.5 | 39.6 | 36.9 | 39.0±3.5 | 41.5 |
| | Extended crust | 5 | 30.8 | 30.5 | 33.7 | 31.7±3.8 | 30.5 |
| | Orogen | 9 | 48.7 | 46.4 | 48.9 | 48.0±6.3 | 46.3 |
| | Bulk CC | -- | 35.7 | 34.8 | 32.7 | 34.4±4.1 | 41.0 |
| | Bulk OC | -- | 7.5 | 7.6 | 8.8 | 8.0±2.7 | -- |
| Mass ($10^{21}$ kg) | Bulk CC | -- | 21.4 | 20.9 | 19.6 | 20.6±2.5 | -- |
| | Bulk OC | -- | 6.3 | 6.4 | 7.4 | 6.7±2.3 | -- |
| | Total Crust | -- | 27.7 | 27.3 | 27.0 | 27.3±4.8 | -- |

[a]The areal percent relative to the total surface of CC based on CRUST 2.0.
[b]The crustal thickness of our RM is the average of three models, and the uncertainty is the surface area weighted average of the half-range uncertainties of all voxels.
[c]CM’95: An study about the average thicknesses of different crustal types by *Christensen and Mooney* [1995].

Table 7: Comparison of HPE concentrations in the continental crust between previous studies and our reference model (**RM**). K, Th and U concentrations are listed as wt. %, µg/g, and µg/g, respectively.

| | | TM[a] | M | W | H | RF | RG | **RM** |
|---|---|---|---|---|---|---|---|---|
| Upper Crust | K | 2.8 | 2.8 | 2.87 | 2.32 | 2.8 | 2.32±0.19 | 2.32±0.19 |
| | Th | 10.7 | 10.7 | 10.3 | 10.5 | 10.7 | 10.5±1.0 | 10.5±1.0 |
| | U | 2.8 | 2.8 | 2.5 | 2.7 | 2.8 | 2.7±0.6 | 2.7±0.6 |
| Middle Crust | K | - | - | - | - | 1.67 | 1.91 | $1.52^{+0.81}_{-0.52}$ |
| | Th | - | - | - | - | 6.1 | 6.5 | $4.86^{+4.30}_{-2.25}$ |
| | U | - | - | - | - | 1.6 | 1.3 | $0.97^{+0.58}_{-0.36}$ |
| Lower Crust | K | 0.28 | 0.53 | 1.31 | 1.24 | 0.50 | 0.50 | $0.65^{+0.34}_{-0.22}$ |
| | Th | 1.06 | 2.0 | 6.6 | 5.6 | 1.2 | 1.2 | $0.96^{+1.18}_{-0.51}$ |
| | U | 0.28 | 0.53 | 0.93 | 0.7 | 0.2 | 0.2 | $0.16^{+0.14}_{-0.07}$ |

[a]Keys to models: TM [*Taylor and McLennan*, 1995]; M [*McLennan*, 2001]; W [*Wedepohl*, 1995]; H [*Hacker et al.*, 2011]; RF [*Rudnick and Fountain*, 1995]; RG [*Rudnick and Gao*, 2003]; **RM** (Reference Model, this study).

Table 8: Comparison of average HPE concentrations, K/U, Th/U and radiogenic heat power in bulk CC between previous studies and our reference model (**RM**).

| | | TM[a] | M | W | H | RF | RG | **RM** |
|---|---|---|---|---|---|---|---|---|
| Bulk CC[c] | K[b] | 1.16 | 1.32 | 1.84 | 1.61 | 1.68 | 1.61 | $1.52^{+0.29}_{-0.22}$ |
| | Th | 4.45 | 5.05 | 7.86 | 7.28 | 6.16 | 6.23 | $5.61^{+1.56}_{-0.89}$ |
| | U | 1.15 | 1.31 | 1.47 | 1.39 | 1.57 | 1.43 | $1.31^{+0.29}_{-0.25}$ |
| | K/U | 10,030 | 10,027 | 12,497 | 11,619 | 10,759 | 11,215 | $11,621^{+3,512}_{-2,516}$ |
| | Th/U | 3.9 | 3.8 | 5.3 | 5.2 | 3.9 | 4.3 | $4.3^{+1.6}_{-1.0}$ |
| | P[d] | 5.6 | 6.3 | 8.5 | 7.9 | 7.7 | 7.4 | $6.8^{+1.4}_{-1.1}$ |

[a]Keys to models are the same as Table 7. [b]Units for HPE concentrations are same as Table 7.
[c]The average HPE concentrations are recalculated based on the same geophysical crustal structure in RM.
[d]P is the radiogenic heat power in TW ($10^{12}$ W) in the bulk CC assuming it has a mass of $20.6\times10^{21}$ kg as RM.

Fig. 1: Schematic drawing of the structure of the reference model (not to scale). Under the continental crust (CC), we distinguish the lithospheric mantle (LM) from depleted mantle (DM), as discussed in Section 2.3. The DM under the CC and the oceanic crust (OC) is assumed to be chemically homogeneous, but with variable thickness because of the depth variation of the Moho discontinuity as well as the continental lithospheric mantle. The boundary between DM and enriched mantle (EM) is determined by assuming that the mass of the enriched reservoir is 17% of the total mantle. The EM is a homogeneous symmetrical shell between the DM and core mantle boundary (CMB).

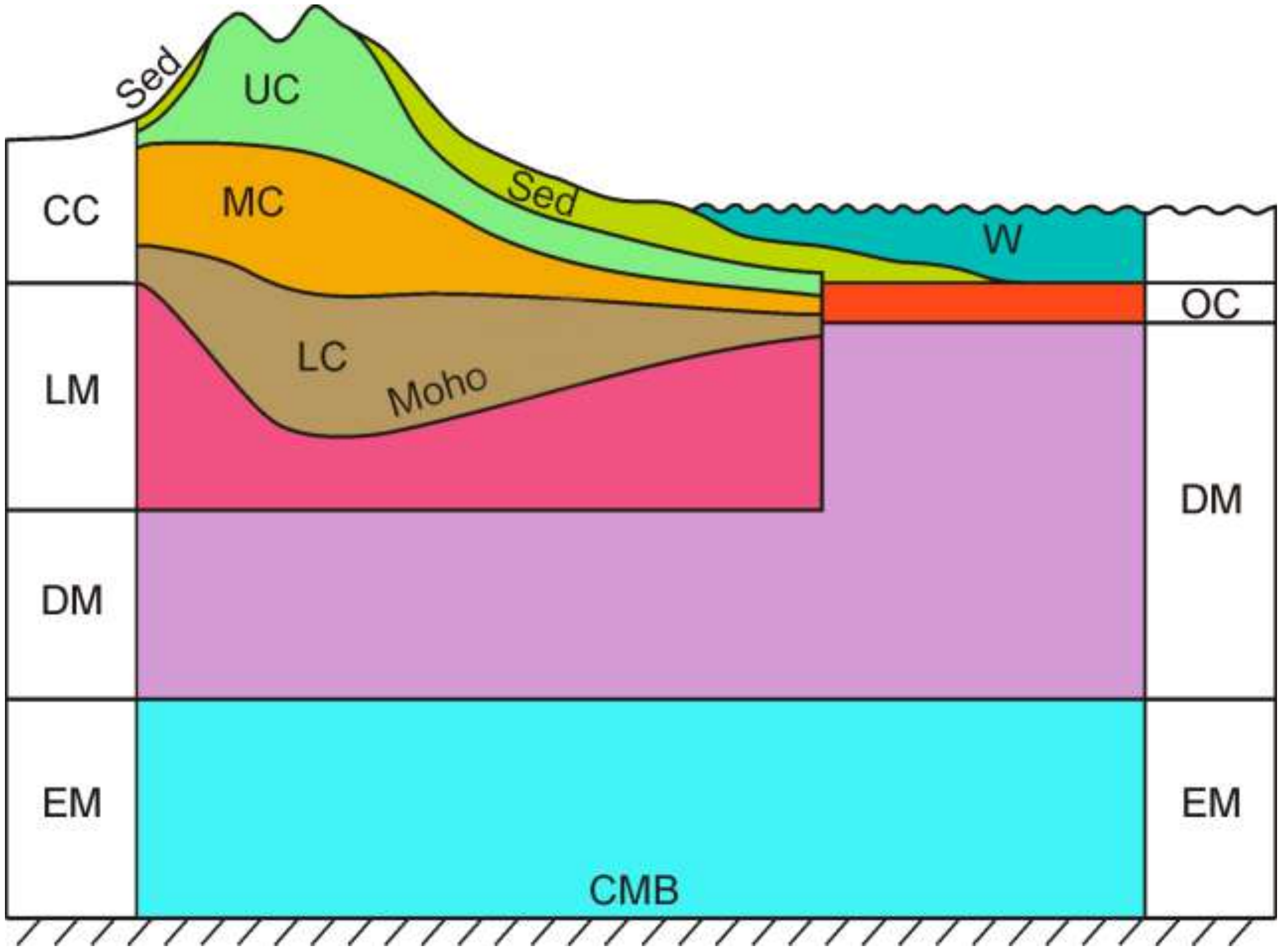

Fig. 2: Distributions of continental crustal thickness (without ice or water) in three global crustal models and our reference model. The average thicknesses of the four models, as shown by the dots lines, are calculated from surface area weighted averaging, and so do not coincide with the mean of the distribution. CRUST 2.0: *Laske et al.* [2001]; CUB2.0: *Shapiro and Ritzwoller* [2002]; GEMMA: *Negretti et al.* [2012]; **RM**: our reference model.

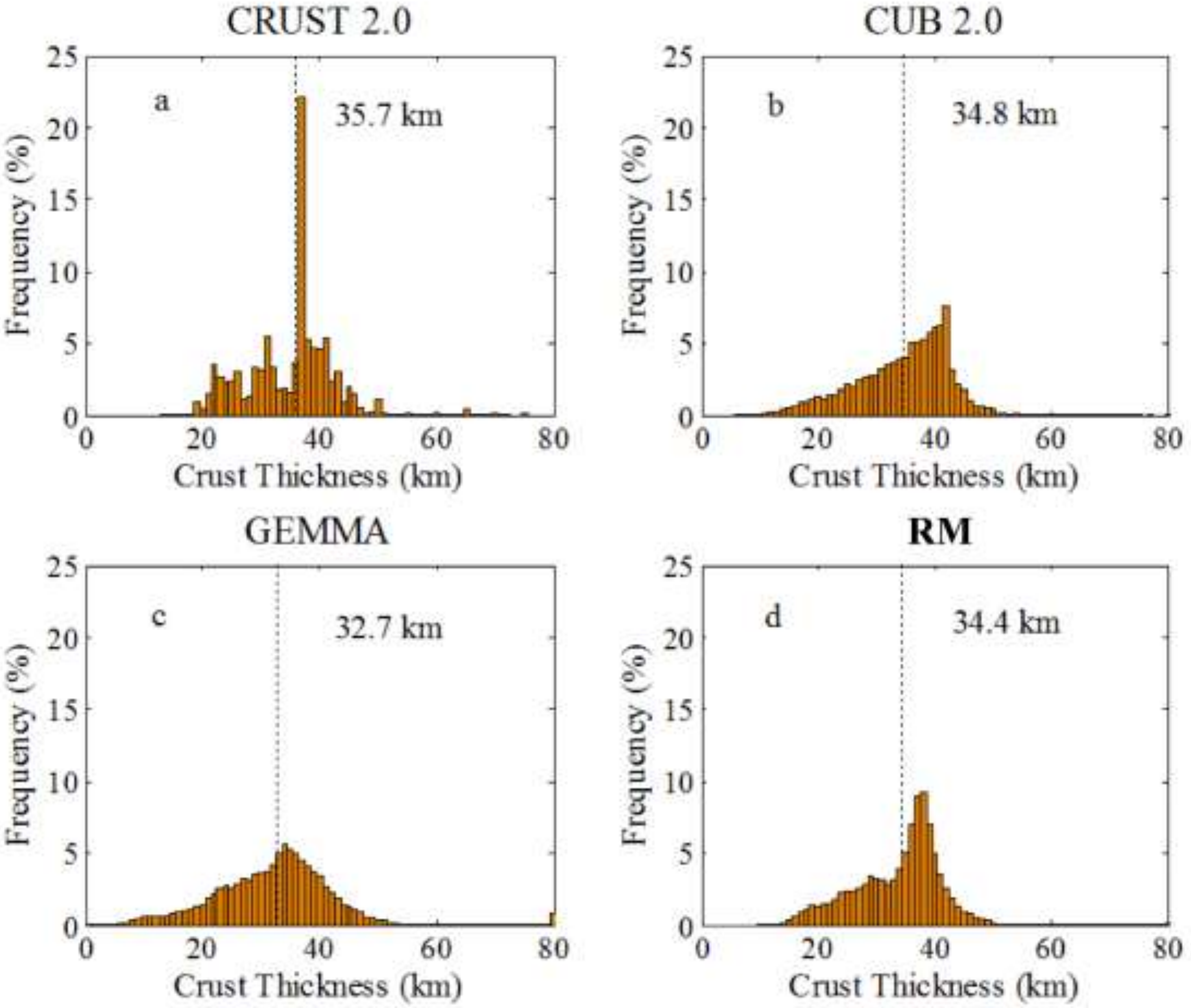

Fig. 3: Laboratory ultrasonic measurements of Vp and Vs for amphibolite facies (open symbols) and granulite facies (closed symbols) meta-igneous rocks versus their $SiO_2$ contents. Felsic rocks are represented by blue diamonds, intermediate rocks by red squares, and mafic rocks by green triangles. Large symbols represent the means of Vp and Vs for felsic, intermediate and mafic rocks, and error bars represent the 1-sigma uncertainties. Vp and Vs generally decrease with increasing $SiO_2$ contents for both amphibolite and granulite facies rocks. This relationship inspires us to estimate the abundances of HPEs in the middle and lower CC using seismic velocity argument.

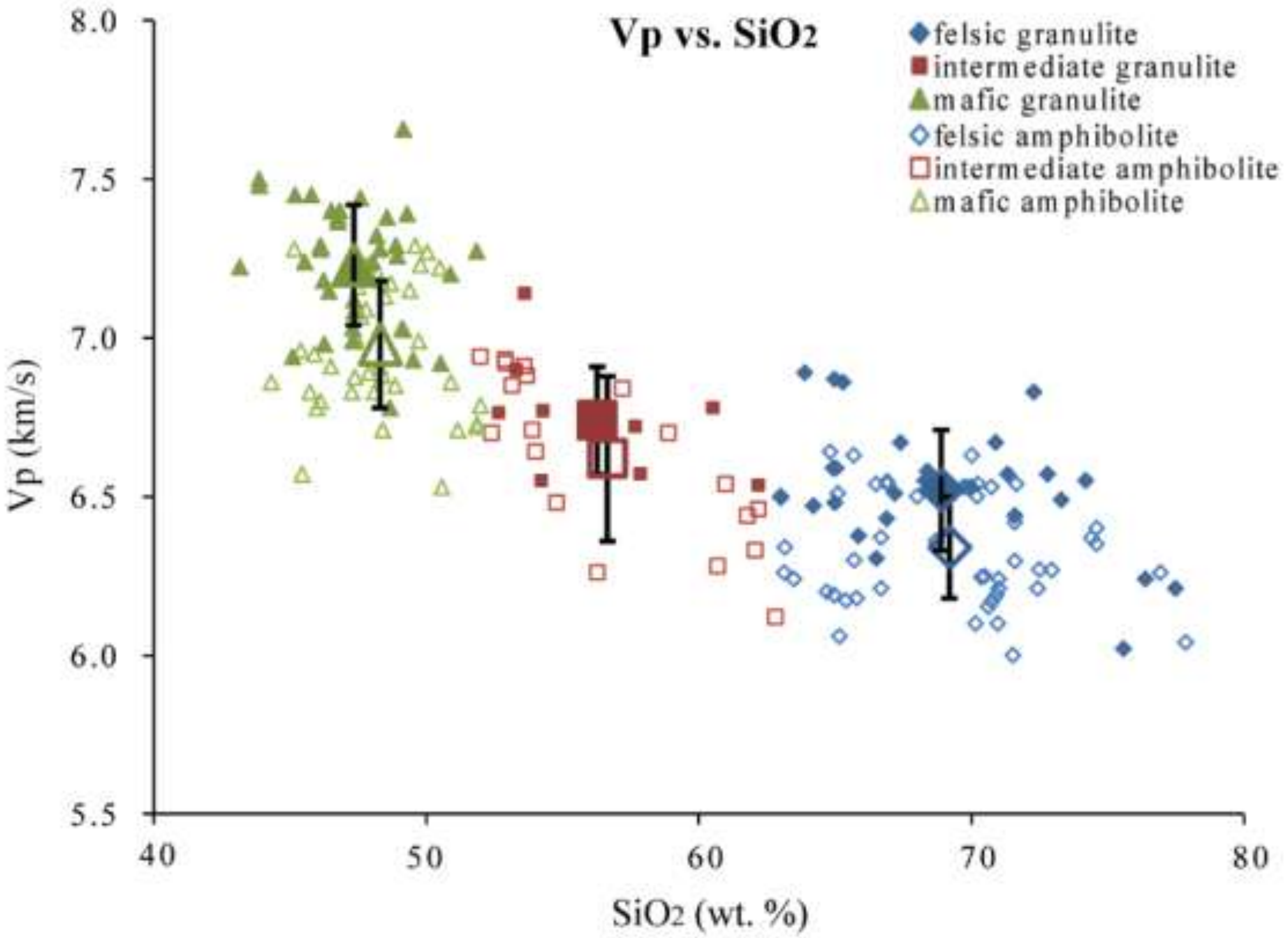

(a)

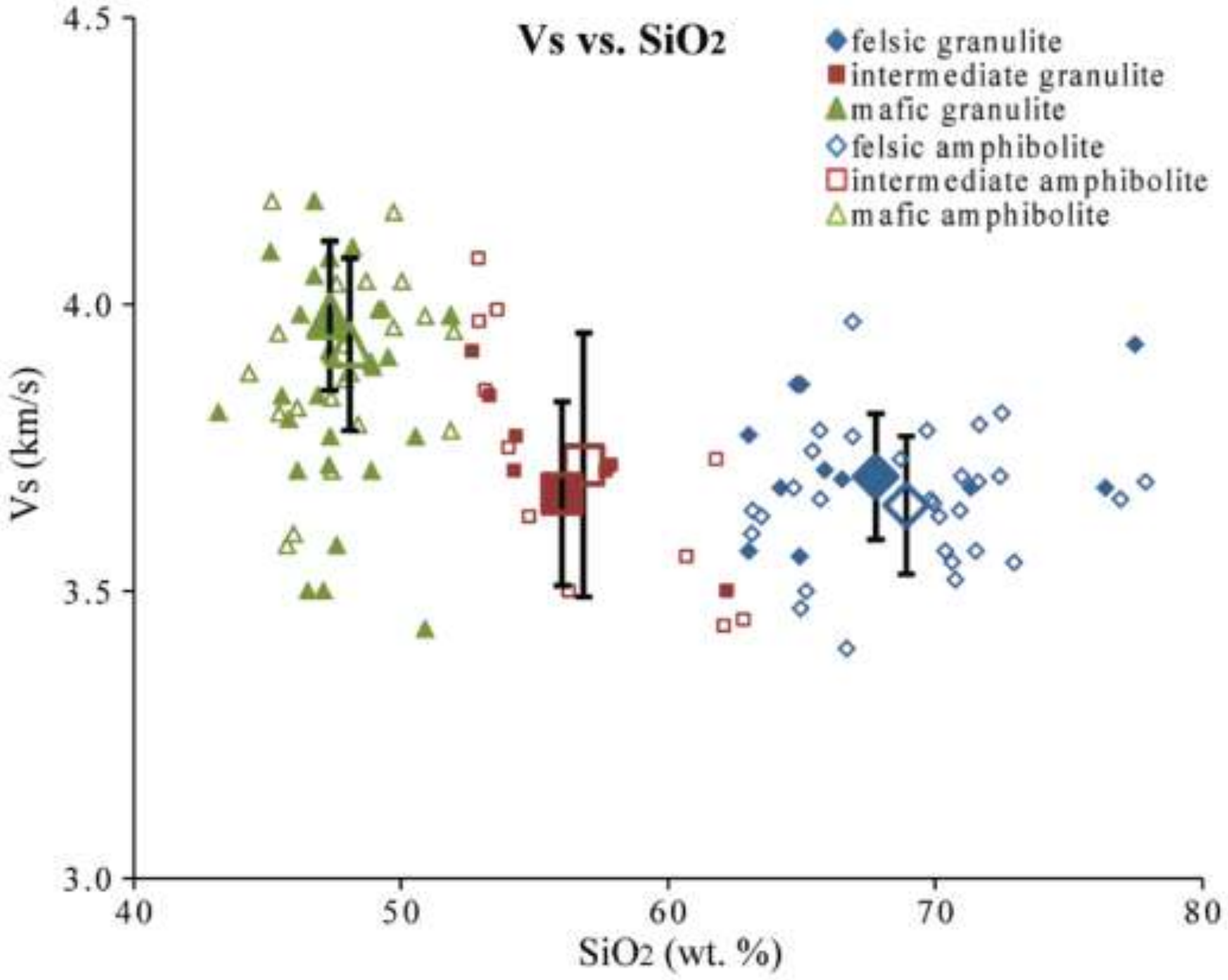

(b)

Fig. 4: Overlapping histograms of laboratory-measured Vp and Vs of felsic (blue) and mafic (red) amphibolite facies (open bars) and granulite facies (filled bars) rocks. The frequency distributions of Vp (a) and Vs (b) of various rock types are generally similar to a Gaussian distribution in character, and the best-fit curves are shown with the histograms.

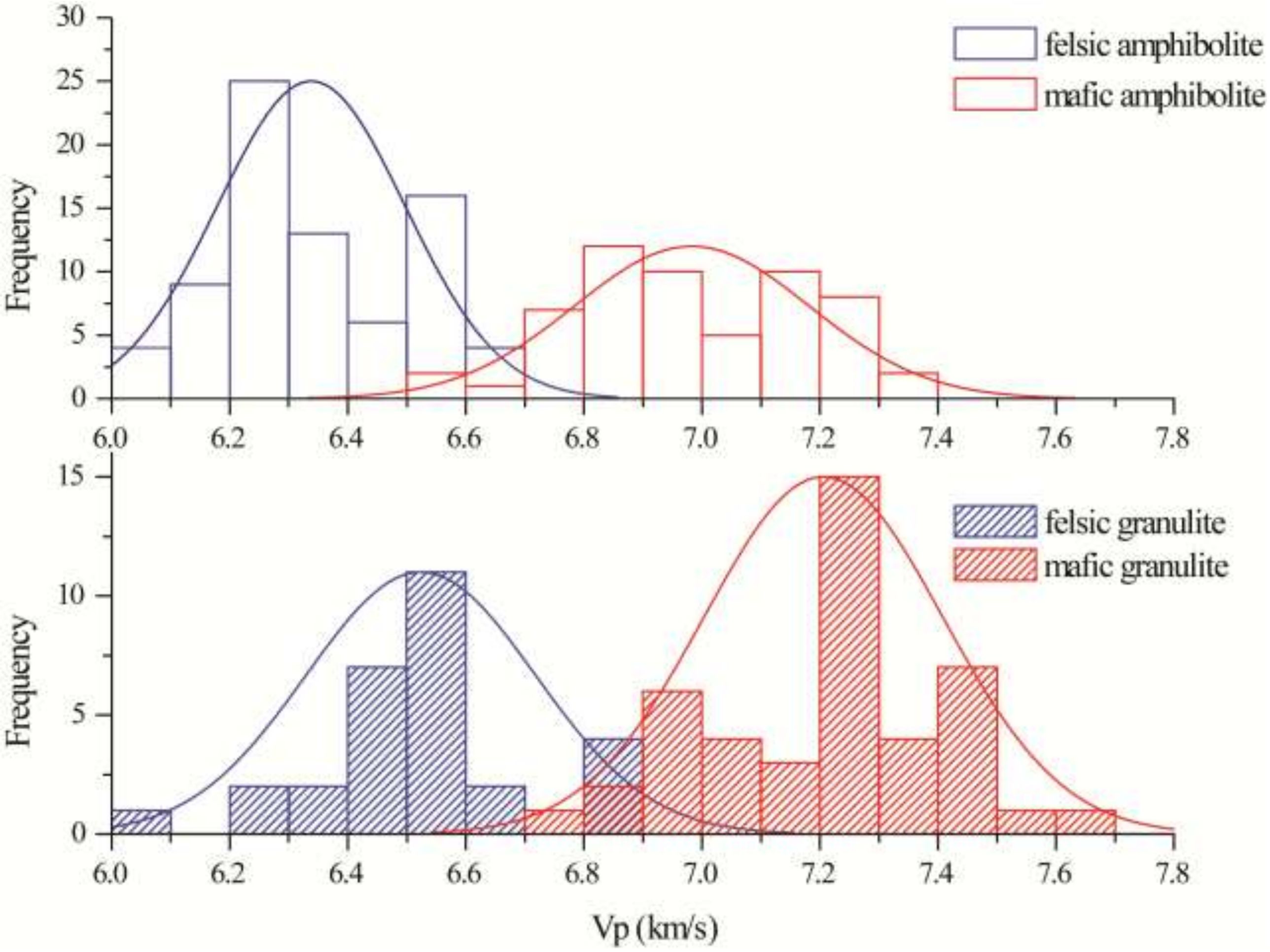

(a)

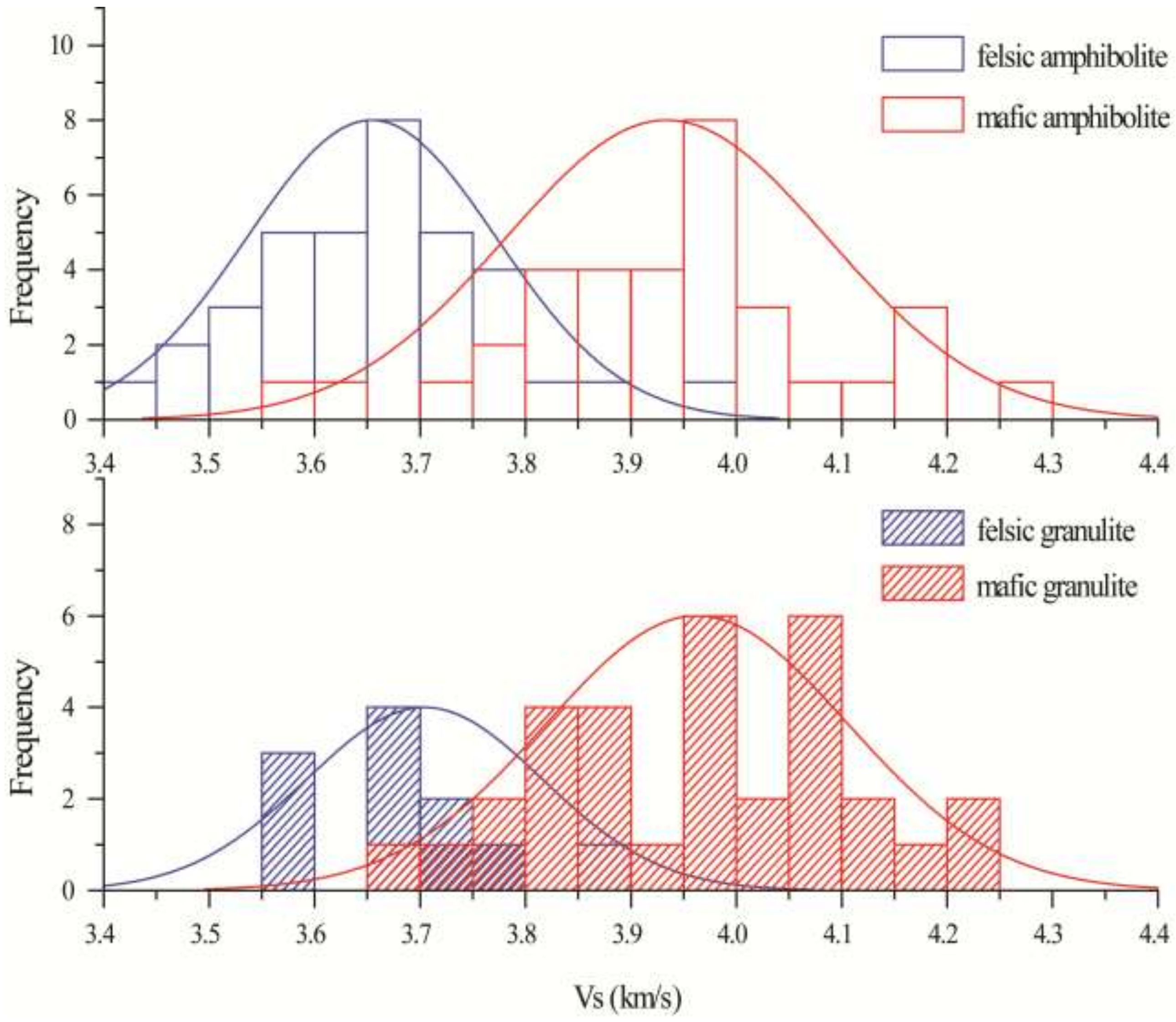

(b)

Fig. 5: Frequency distributions of U abundances of felsic and mafic amphibolite facies rocks, after applying the 1.15 sigma filter as discussed in Section 4.3.1, are strongly positively skewed. Taking the logarithm of the abundances converts the distributions to a more Gaussian geometry. Th and K abundances in both amphibolite and granulite facies rocks show the same characteristics.

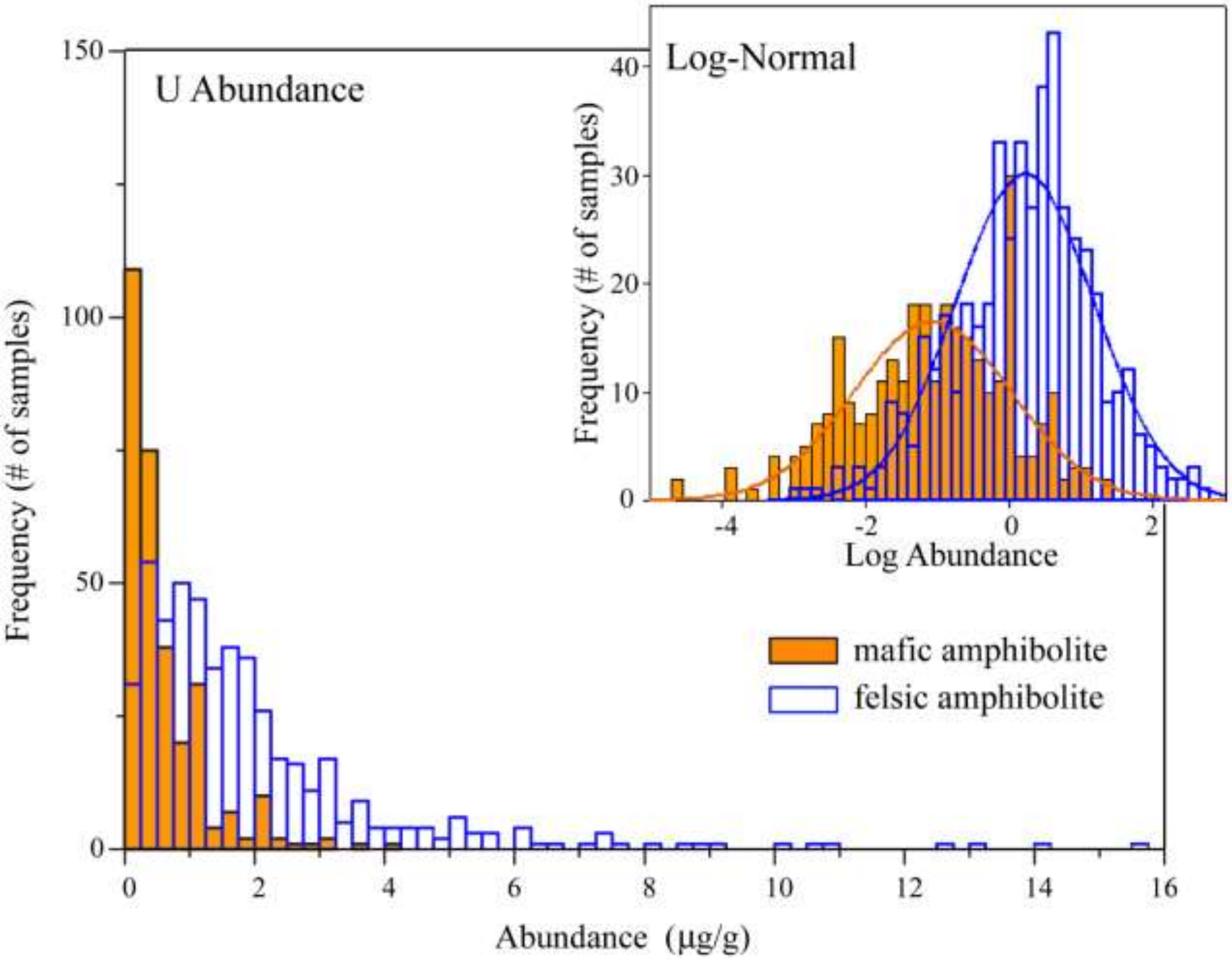

Fig. 6: Box-and-whisker diagram showing the HPE abundance dispersion in the amphibolite and granulite facies rocks after filtering. The numbers of samples are shown above or below the whiskers. The lines near the center of each box represent the median values. The bottom and top edges of the box are the 25th and 75th percentile, respectively (also known as the lower and upper quartiles). The difference between the lower and upper quartile is referred to as the interquartile range (IQR). The high whisker represents the boundary within 1.5 IQR above the upper quartile; the lower whisker represents either the minimum value of the data distribution or the boundary within 1.5 IQR below the lower quartile. Any data that are not included within the whiskers are plotted as outliers (crosses).

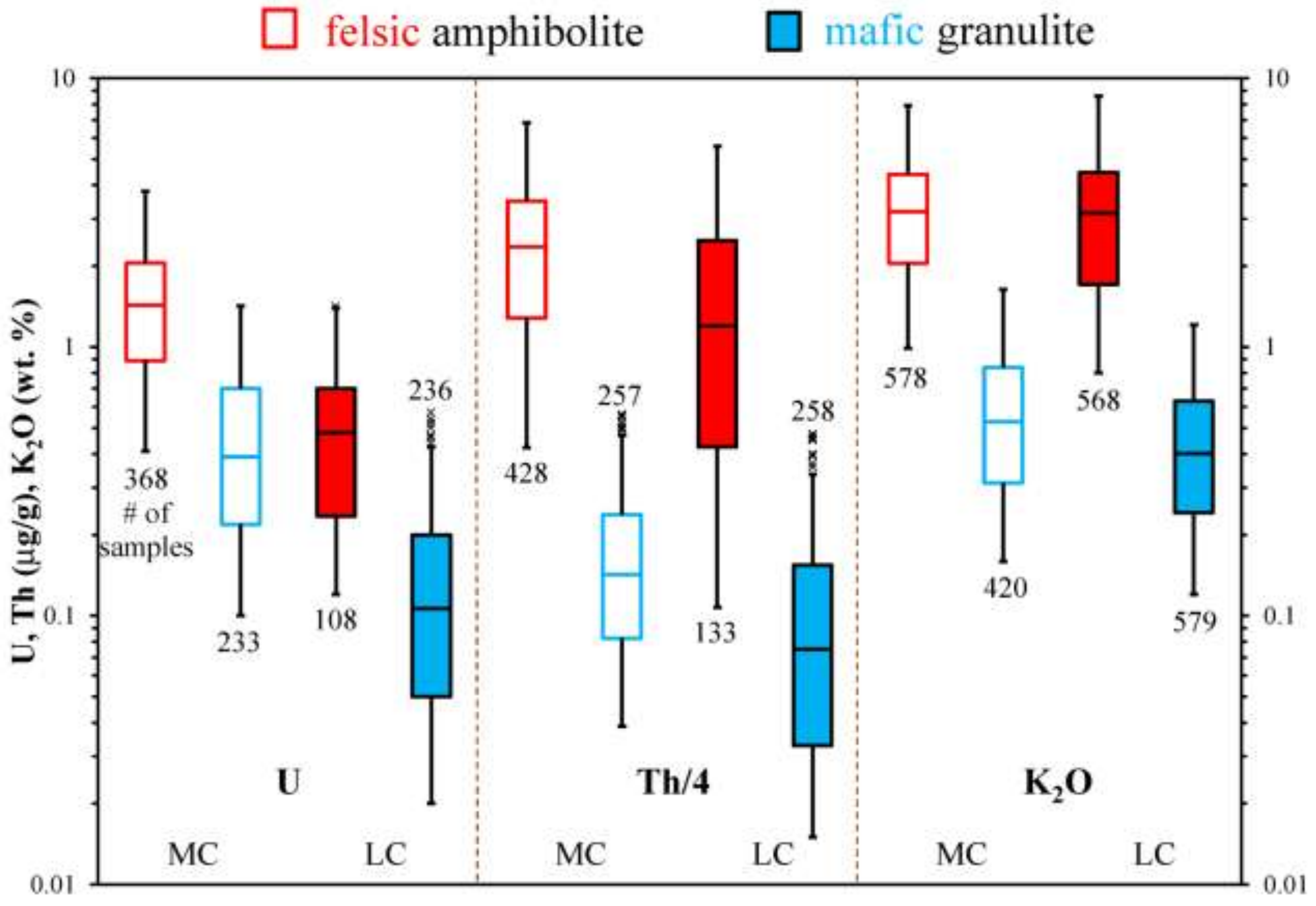

Fig.7: Thickness of crust (a) and its relative uncertainty (b) of our reference model.

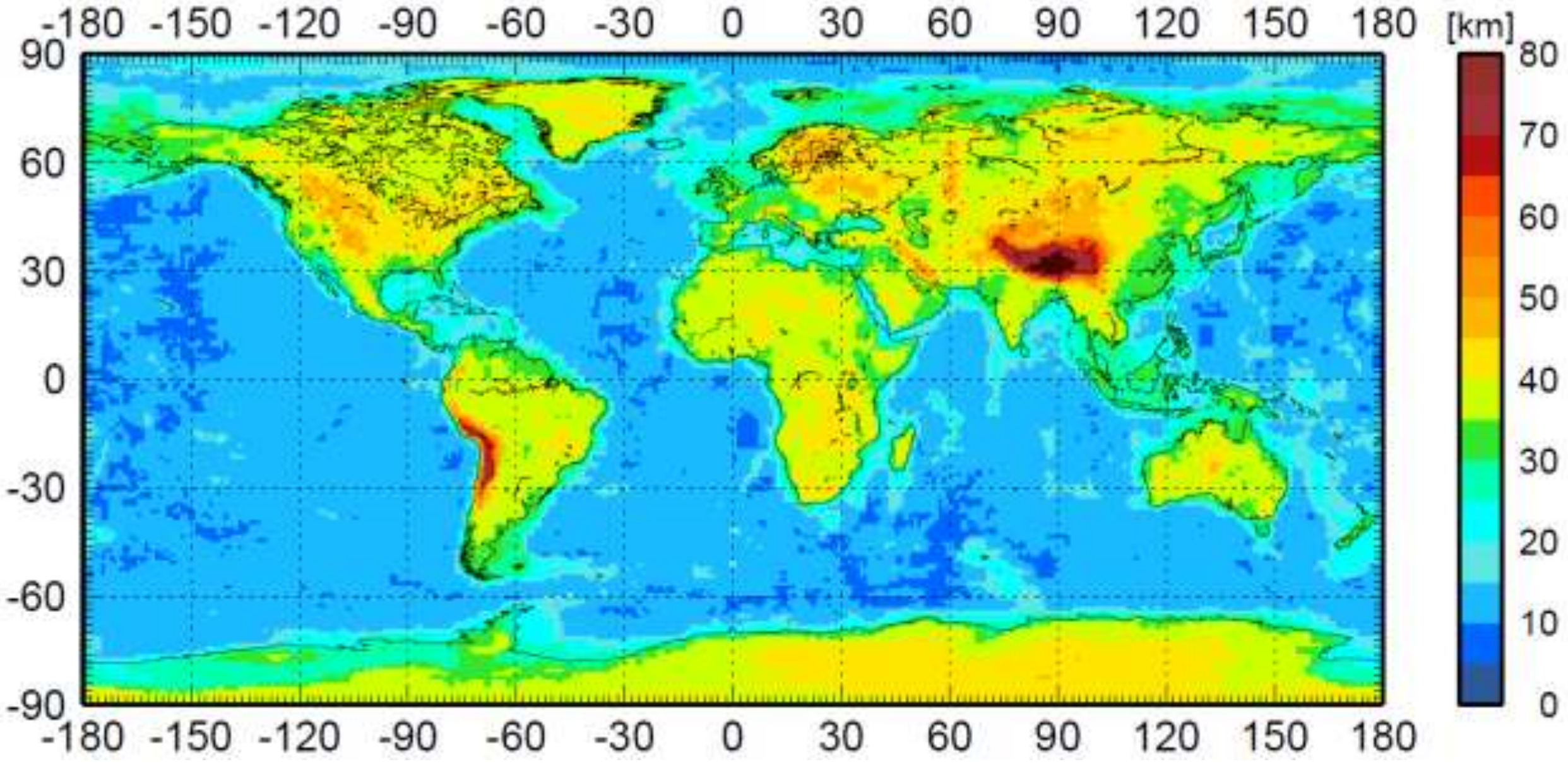

(a)

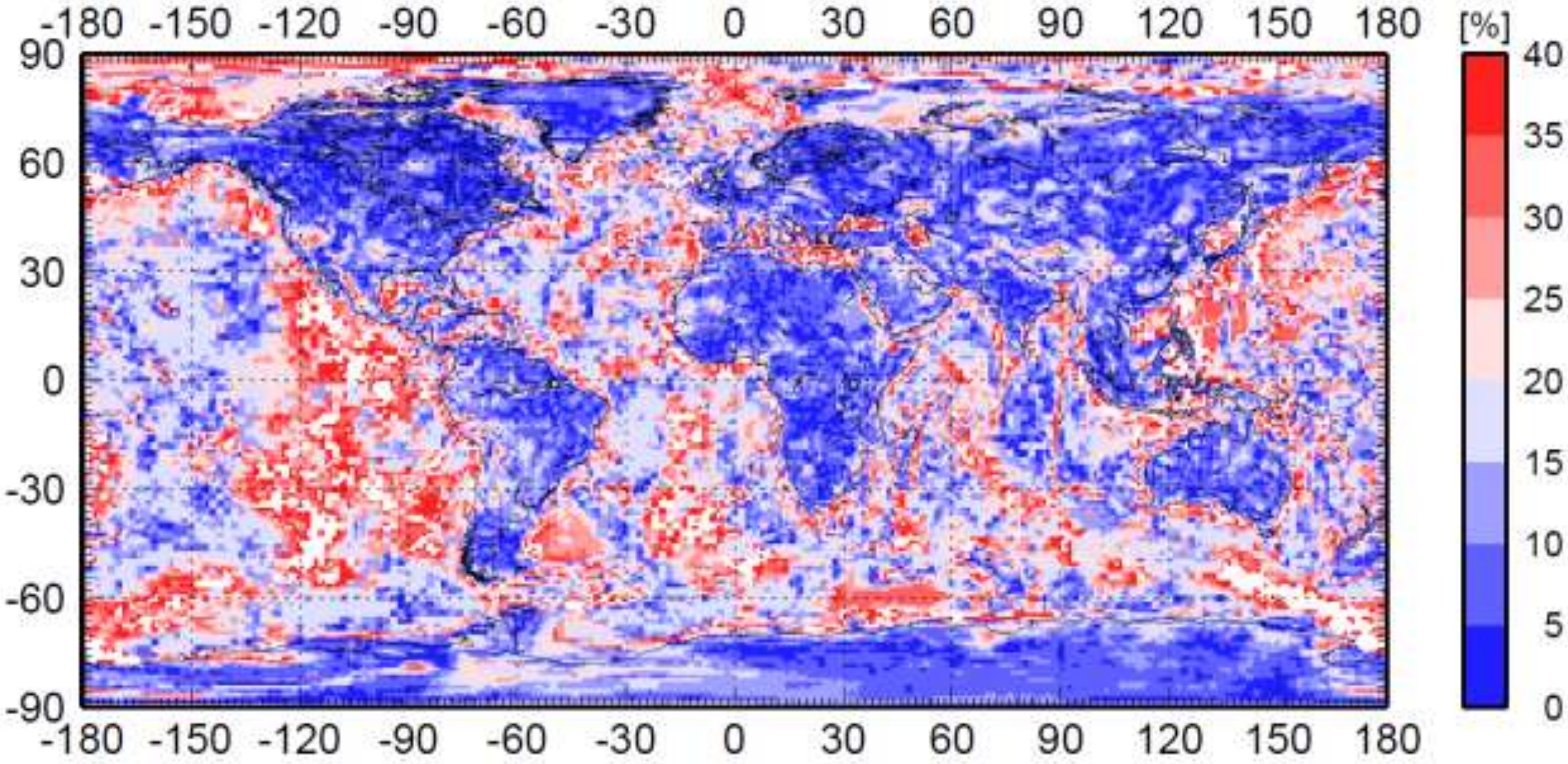

(b)

Fig. 8: The abundance of U in the middle (a) and lower CC (b) using seismic velocity argument.

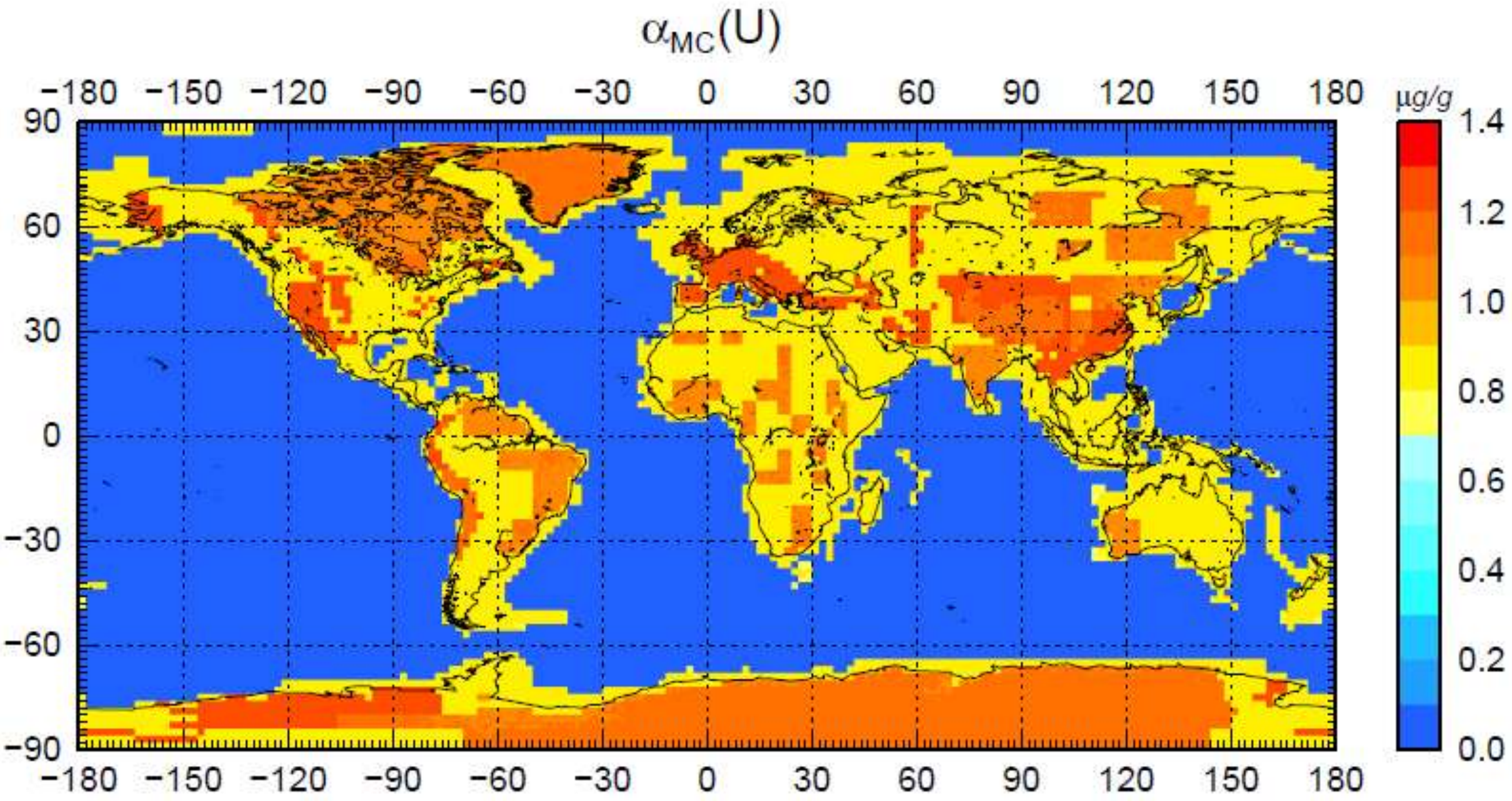

(a)

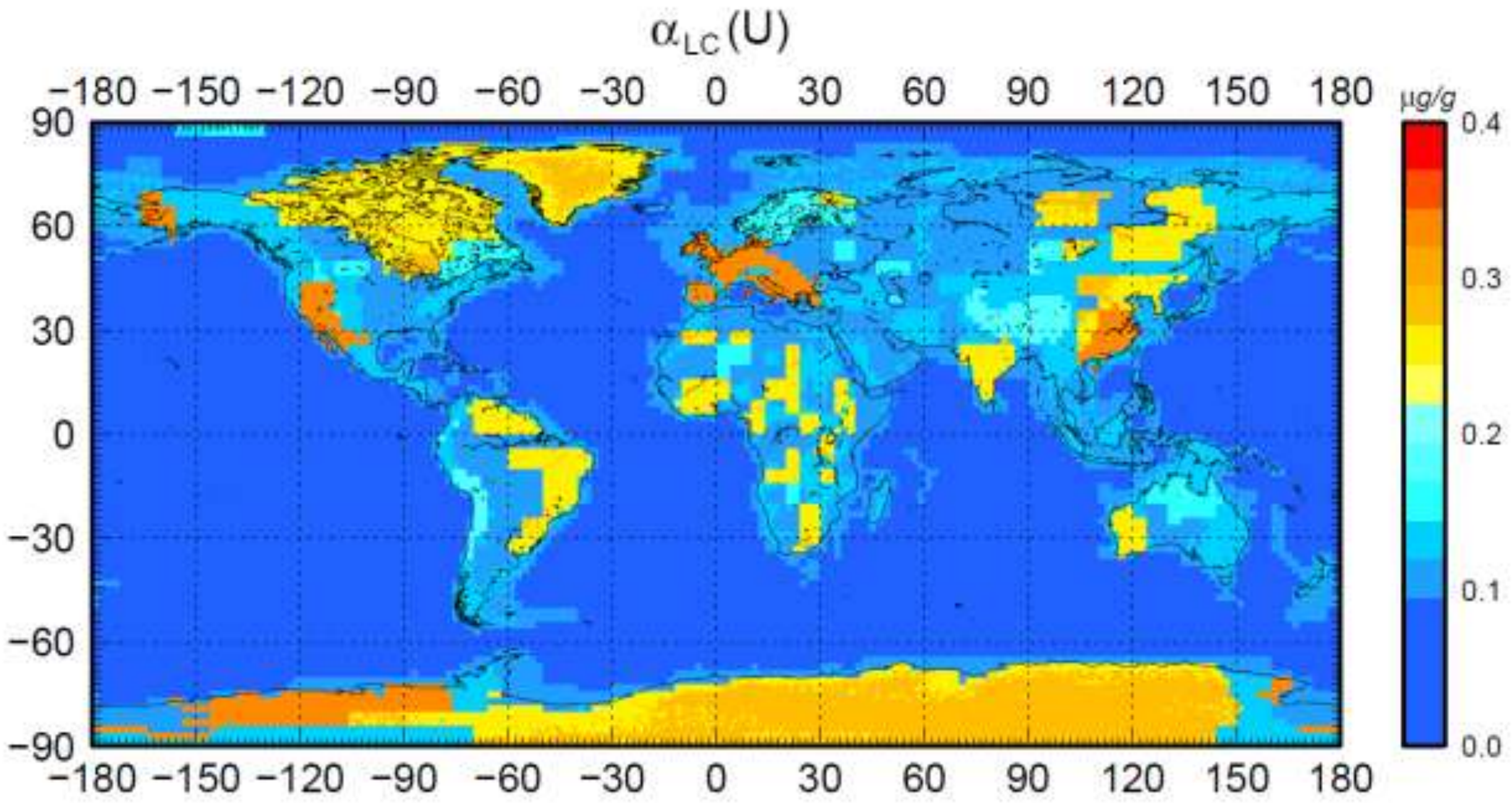

(b)

Fig. 9: Geoneutrino signal at Earth's surface. The unit is Terrestrial Neutrino Unit (TNU) as discussed in Section 5.

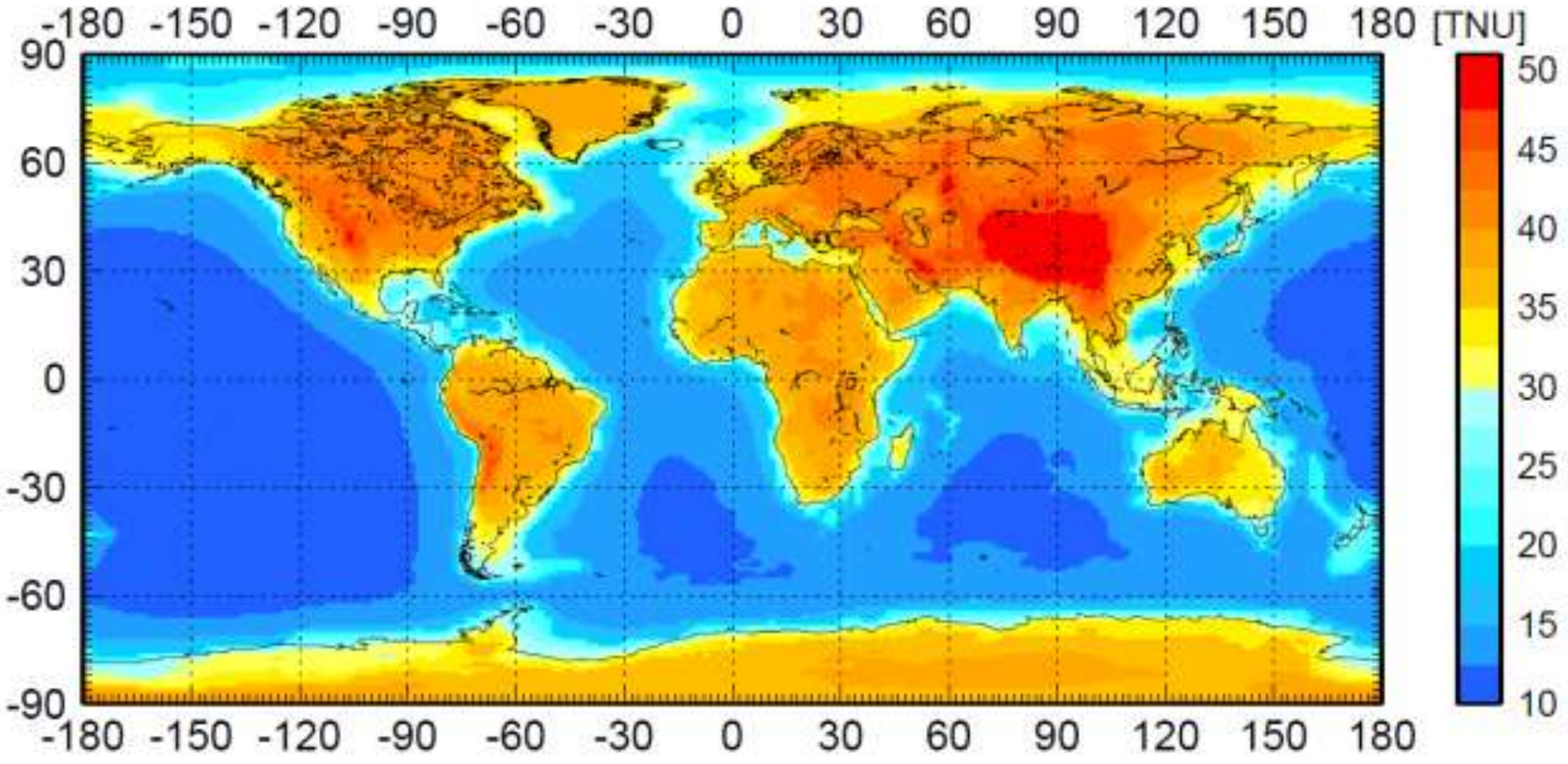

Fig. 10: Predicted geoneutrino signals from the mantle (DM+EM; blue) and overlaying lithosphere (crust+CLM; yellow to red) for 16 geographic locations.

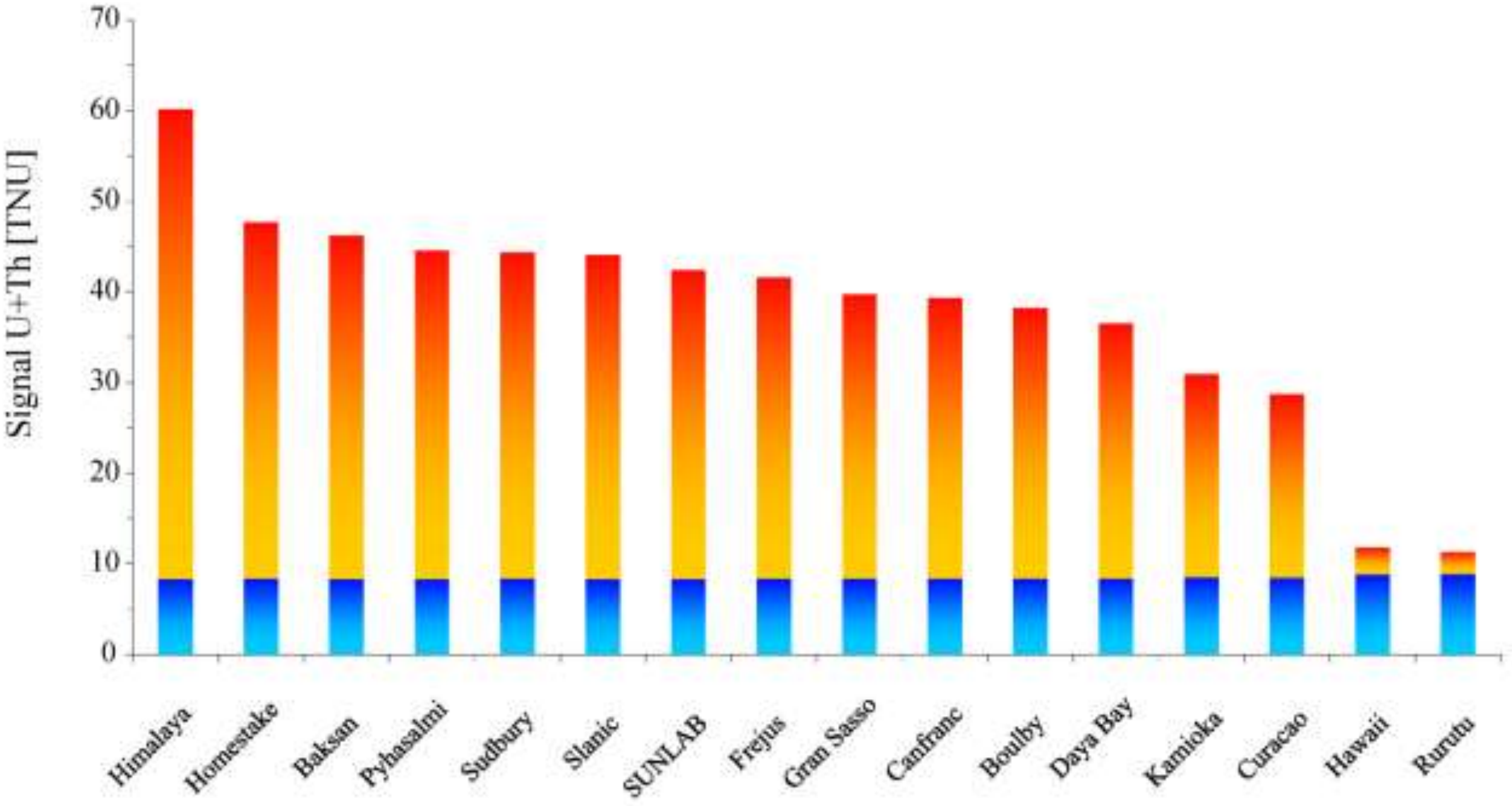

# Supplement Material

## 1. Descriptive statistics

Descriptive statistics is the discipline of quantitatively describing the main characteristics of a data set. The measures involved in describing a collection of data are measures of **central tendency** and measures of **variability or dispersion**. Statistically, the central tendency describes the way in which quantitative data tend to cluster around some “central value”.

### *1.1 Central limit theorem for independent samples*

Let $\{X_1, ..., X_n\}$ be a random sample of size n that is a sequence of independent and identically distributed random variables drawn from a population of expected value μ and finite variances $\sigma^2$. The sample average (arithmetic mean):

$$S_n = \frac{X_1 + X_2 + ...X_n}{n} \qquad \text{(eq. S1)}$$

of these random variables converges to the expected value μ as n → ∞. As n gets larger, the distribution of the difference between the sample average $S_n$ and the expected value μ of the population:

$$\lim_{n \to +\infty} \sqrt{n}\left(S_n - \mu\right) = 0 \qquad \text{(eq. S2)}$$

approximates the normal distribution, with mean 0 and variance $\sigma^2$.

The central limit theorem is the statistical basis for using arithmetic mean of the samples to represent the expected value of a normally distributed population from which these samples are randomly drawn. However, arithmetic mean is not a robust statistic, meaning that it is greatly influenced by outliers. Notably, for skewed distributions, more robust statistics such as the median, rather than arithmetic mean, may be a better description of central tendency.

Despite the fact that there are many different measures of the central tendency of samples, the arithmetic mean is equal to the median, mode, and other measures of the central tendency of a normal (Gaussian) distribution.

### *1.2 Log-normal (geometric) mean and log-normal distribution*

A log-normal distribution is a continuous probability distribution of a random variable whose natural logarithm is normally distributed (N $(\mu, \sigma^2)$). Therefore, the arithmetic mean of the natural logarithm of a log-normal distribution best describes its central tendency, and the log-normal mean (geometric mean) of the log-normal distribution is $e^{\mu}$. Because the natural logarithim of a log-normal variable is symmetric and quantiles are preserved under monotonic transformations, the geometric mean of a log-normal distribution is equal to its median. The 1-sigma uncertainty for the geometric mean of a log-normal distribution is $^{+(e^{\mu+\sigma}-e^{\mu})}_{-(e^{\mu}-e^{\mu-\sigma})}$.

### *1.3 HPE abundances in this study*

The distributions of U, Th and K abundances in the rocks are best-fit by a log-normal, but not normal, distribution (see Fig. 5 and 6 in the text). In this case the central limit theorem is not valid, and the arithmetic mean of compiled samples loses its property of best measure of central tendency because it is greatly influenced by outliers. Therefore, geometric means and associated uncertainties are employed to describe the measures of central tendency and dispersion of the collected data.

## 2. Beyond descriptive statistics

The reported average abundances of HPEs in the deep crust and continental lithospheric mantle in this paper are the log-normal (geometric) means of the distributions, and they serve the purpose of measuring the central tendencies (most probable HPE abundances) of these distributions. A potential confusion may be present when treating geometric mean in the same way as arithmetic mean beyond the descriptive statistics.

Arithmetic mean always maintains the feature that the sum of collected data is the product of arithmetic mean and the size of the data set. Other measures of central tendency, such as geometric mean, of a distribution are not defined by eq. S1, and thus do not simply yield information about the sum of the data as arithmetic mean. When dealing with significantly skewed distributions, their probability density functions can be described through Monte Carlo simulation (Fig. S1), in preference to measures of central tendency, such as the log-normal mean. The basic idea of MC simulation is to generate a large number of random samples that follow the characteristics (central tendency and dispersion) of input distributions, and the distribution of output is obtained for further statistical description.

An example of calculating the mass of U in the middle continental crust is helpful for understanding the MC simulation. We assume that the mass of middle continental crust is $m \pm \sigma_m$ (Gaussian distribution) and that the abundance of U in this reservoir is $\exp(N(\mu, \sigma^2))$ (log-normal distribution). To calculate the mass of U, simply multiplying the abundance of U ($e^{\mu}$) by the mass of the reservoir (m) is not the right approach. In a MC simulation, a large number of random samples are generated that follow the two distributions (Fig. S1), and the skewed distribution of the mass of U is obtained by multiplying the generated samples for mass of middle continental crust and abundance of U. Further descriptive statistics are required to analyse the resulting skewed distribution (median value and 68% population are recommended for describing the central value and 1-sigma uncertainty of such a skewed distribution).

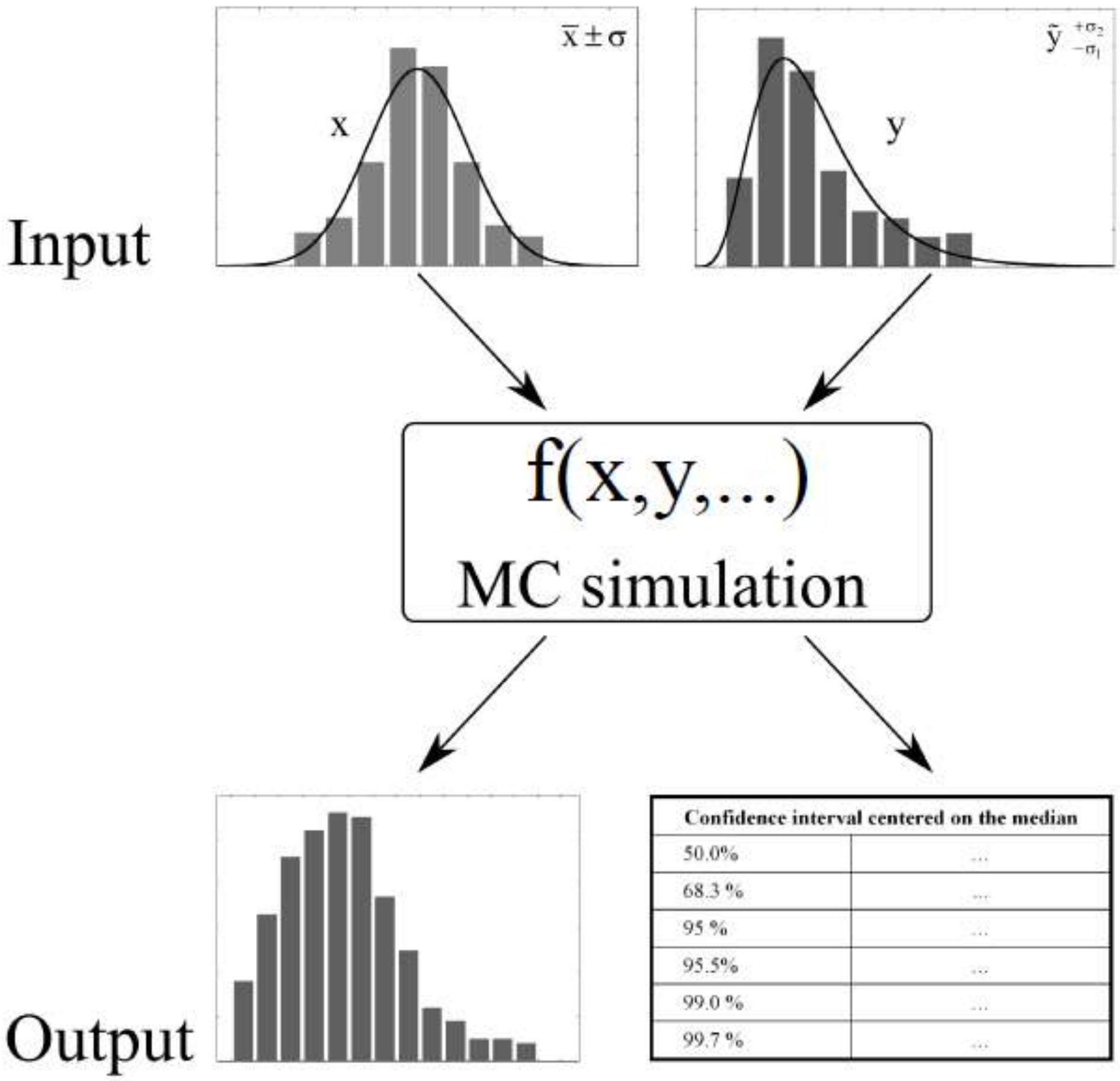

Fig. S1: Illustration of Monte Carlo simulation. See also Section 5 in the text for full explanation of how to perform MC simulation.